\documentclass[%
 aip,
 amsmath,amssymb,
reprint,%
]{revtex4-1}

\usepackage{graphicx}
\usepackage{dcolumn}
\usepackage{bm}

\usepackage[utf8]{inputenc}
\usepackage[T1]{fontenc}
\usepackage{mathptmx}
\usepackage{comment}
\usepackage{xcolor}

\begin{document}


\title[ ]{Meso-resolved simulations of shock-to-detonation transition in\\nitromethane with air-filled cavities}

\author{X.C. Mi} \email{xcm20@cam.ac.uk; xiaocheng.mi@mail.mcgill.ca}
\affiliation{Cavendish Laboratory, Department of Physics, University of Cambridge, Cambridge, United Kingdom CB3 0HE}
 \affiliation{Department of Mechanical Engineering, McGill University, Montreal, Quebec, Canada H3A 0C3}
\author{L. Michael}
\author{E. Ioannou}%
\author{N. Nikiforakis}%
\affiliation{Cavendish Laboratory, Department of Physics, University of Cambridge, Cambridge, United Kingdom CB3 0HE}

\author{A.J. Higgins}
\affiliation{%
Department of Mechanical Engineering, McGill University, Montreal, Quebec, Canada H3A 0C3
}%

\author{H.D. Ng}
\affiliation{%
Department of Mechanical, Industrial and Aerospace Engineering, Concordia University, Montreal, Quebec, Canada H3G 1M8
}%

\date{\today}

\begin{abstract}
Two-dimensional, meso-resolved numerical simulations are performed to investigate the complete shock-to-detonation transition (SDT) process in a mixture of liquid nitromethane (NM) and air-filled, circular cavities. The shock-induced initiation behaviors resulting from the cases with neat NM, NM with an array of regularly spaced cavities, and NM with randomly distributed cavities are examined. For the case with randomly distributed cavities, hundreds  of cavities are explicitly resolved in the simulations using a diffuse-interface approach to treat two immiscible fluids and GPU-enabled parallel computing. Without invoking any empirically calibrated, phenomenological models, the reaction rate in the simulations is governed by Arrhenius kinetics. For the cases with neat NM, the resulting SDT process features a superdetonation that evolves from a thermal explosion after a delay  following the passage of the incident shock wave and eventually catches up with the leading shock front. For the cases wherein mesoscale heterogeneities are explicitly considered,  a gradual SDT process is captured. These two distinct initiation behaviors for neat NM and heterogeneous NM mixtures agree with experimental findings. Via examining the global reaction rate of the mixture, a unique time scale characterizing the SDT process, i.e., the overtake time, is measured for each simulation. For an input shock pressure less than approximately $9.4~\mathrm{GPa}$, the overtake time resulting from a heterogeneous mixture is shorter than that for neat NM. This sensitizing effect is more pronounced for lower input shock pressures. A random distribution of cavities is found to be more effective in enhancing the SDT process than a regular array of cavities. Statistical analysis on the meso-resolved simulation data provides more insights into the mechanism of energy release underlying the SDT process. Possible directions towards a quantitatively better agreement between the experimental and meso-resolved simulation results are discussed. 

\end{abstract}

\maketitle

\section{\label{sec1}Introduction}

Condensed-phase explosives can be significantly sensitized to shock initiation by the introduction of mesoscale heterogeneities. This sensitizing effect is attributed to the formation and growth of localized high-temperature regions, or the so-called ``hot spots'', as an incident shock wave interacts with the heterogeneities. Since the concept of hot-spot sensitization was first proposed by Bowden and Yoffe in the 1940s~\cite{BowdenYoffe1948hot}, a substantial body of research has been carried out to elucidate the underlying mechanisms. 

In contrast to most solid and slurry explosives with intrinsic heterogeneities that are highly irregular in shape, size, and spatial distribution, the size and number density ($\rho_\mathrm{N}$) of heterogeneities added into a liquid explosive matrix is relatively controllable. Taking advantage of this feature, the use of liquid nitromethane (NM) mixed with micron-sized solid particles or hollow spheres has been explored to experimentally probe the sensitizing effect of mesoscale heterogeneities on shock-to-detonation transition (SDT) in high explosives. Campbell~\textit{et al}. reported in 1961 that the build-up of a detonation in NM with trapped bubbles of gas or pieces of solid is significantly faster than that in neat NM.~\cite{Campbell1961Liquid} Later, via \textit{in-situ} measurement of particle velocity, it has been revealed that the SDT process in NM sensitized by the addition of silica beads or glass micro-balloons (GMBs) features a reaction front growing in the vicinity of the incident shock~\cite{Sheffield1989Report}, in contrast to the SDT in neat NM wherein the reaction front is initiated significantly behind the incident shock, evolves into a superdetonation, and eventually overtakes the incident shock.\cite{Chaiken1960,Campbell1961Liquid} To further interpret these macroscopic measurements, one needs to gain direct insights into the mechanisms of how hot spots are formed, evolve, and collectively contribute to the energy release process. Thus, studies focused on illustrating the detailed wave dynamics arising from the interaction of an incident shock wave with mesoscale heterogeneities have been motivated.

The wave and interface morphologies evolving over the course of shock-induced collapse of cavities have been experimentally observed using high-speed photography.~\cite{DearField1988,Dear1988Nature,Field1992,Field1992hot,BourneField1992,Bourne1999,Bourne1999_react,Bourne2003,Swantek2010AIAA,Swantek2010JFM} Wave interactions resulting from the collapse of neighboring cavities have been revealed by Dear and Field and later researchers.\cite{DearField1988,Bourne1999,Bourne1999,Swantek2010JFM} To enable visualization, millimeter-sized or larger cavities were considered in these experimental studies while typical GMBs added into liquid NM, for example, are of diameters on the order of $10$-$100$~$\mu\mathrm{m}$. It is rather challenging, although attempted by Bourne and Milne~\cite{Bourne2003}, to experimentally resolve the temperature field arising from the collapse on the spatial scale of a single cavity or average spacing between cavities. Therefore, to understand the formation and evolution of hot spots, shock-induced collapse of micron-sized cavities has been otherwise investigated via numerical simulations.

One- and two-dimensional simulations of a cavity subjected to a planar incident shock wave were performed in the 1960s. As only inert scenarios were considered by Evans~\textit{et al}.~\cite{Evans1962}, Mader, for the first time, demonstrated how the formation of hot spots triggers a rapid burnout of the reactive material surrounding a cavity or a solid pellet~\cite{Mader1963,Mader1965}. Three-dimensional simulations of shock-induced collapse of an array of $91$ regularly staggered, micron-sized cavities in NM were performed by Mader in the 1980s.\cite{Mader1984} The simulation results showed that hot spots of increasingly greater temperature and size are formed as the leading shock passes across layers of cavities. Recent computational effort has been dedicated to revealing finer detail of the resulting temperature field from shock--cavity interactions.\cite{Ball2000,Swantek2010AIAA,Swantek2010JFM,Hawker2012,Lauer2012,Ozlem2012,Michael2014DS,Kapila2015,Betney2015,Apazidis2016,Michael2018_I,Michael2018_II,Michael2019Book} Two- and three-dimensional simulations of the collapse of a single air-filled cavity in liquid NM have been performed by Michael and Nikiforakis.\cite{Michael2018_I,Michael2018_II} These authors demonstrated that hot spots are not only formed due to the penetration of a micro-jet along the centerline of the cavity, but also generated behind Mach stems occurring away from the centerline. These simulation results indicate that hot spots are of irregular shapes and undergo complex evolutions. Thus, it is rather hard to measure the size of a hot spot, and it might not be informative to characterize a hot spot by its temperature without assessing the mass of reactive material associated with it. After revealing detail on mesoscales, how can one computationally examine the mechanism of a large number of hot spots affecting the SDT process in an energetic medium? Two strategies prevail nowadays: Meso-informed and meso-resolved simulations.

In meso-informed simulations, an explosive mixture with mesoscale heterogeneities is treated as a homogeneous medium with a phenomenological (or so-called ``meso-informed'') model---the Ignition and Growth (IG) model proposed by Lee and Tarver~\cite{LeeTarver1980} is the most commonly adopted---that governs the rate of energy release for the given local flow and thermodynamic properties. The computational studies using a variety of phenomenological reaction models have recently been reviewed in detail by Handley \text{et al}.\cite{HandleyReview2018} Conventional strategies to develop a meso-informed reaction model consists of, first, constructing the ignition and growth rate terms based on some micro-mechanistic models for single-cavity collapse, and then calibrating the values of the governing parameters against empirical data. Novel approaches using machine-trained surrogate models for the IG rates have been recently applied by Sen~\textit{et al}. to simulate SDT processes in HMX (cyclotetramethylene-tetranitramine, also called octogen).~\cite{Sen2018JAP} These surrogate models are trained by ensembles of simulation data of single-cavity collapse under various conditions (e.g., incident shock strength, pulse duration, cavity size and shape, etc.). Such simulations, however, are unlikely able to adequately describe the mutual influence among a large number of randomly distributed hot spots. Hence, meso-resolved simulations might be a more pertinent approach to examine the collective effect of hot spots.

In meso-resolved simulations, a statistically significant amount of mesoscale heterogeneities is explicitly considered; instead of invoking a phenomenological model, the reaction rate is determined by a model based on detailed or reduced chemical kinetics, i.e., activated kinetics in an Arrhenius form. With advanced computing technology, to perform meso-resolved simulations of detonation phenomena in high explosives has recently become feasible. Via performing meso-resolved simulations, Rai~\textit{et al}.~\cite{Rai2015JAP} and Kim~\textit{et al}.~\cite{Kim2018JMPS} demonstrated that the characteristics of mesoscale features determine the macroscopic ignition behavior of HMX and polymer-bonded explosives (PBXs), respectively. These simulations, however, have not yet been extended to time or length scales over which a complete SDT process occurs.

In the current study, two-dimensional, meso-resolved simulations of complete SDT processes in liquid NM mixed with air-filled cavities are performed. The resulting initiation behaviors from the cases with randomly distributed cavities are compared to those from the cases with a regular array of cavities and neat NM. These computations are made amenable through the implementation of a diffuse-interface approach to treat two immiscible fluids~\cite{Michael2016JCP} and a CUDA-based (CUDA is an acronym for Compute Unified Device Architecture) parallel computing on general-purpose graphic processing units (GPGPUs). The objective of this work is to explore the capability of meso-resolved simulations in capturing the experimentally observed SDT behaviors of GMB-sensitized NM without invoking a calibrated IG reaction rate model. Via performing statistical analysis on the full information of the resulting flow field, a further elucidation of the hot-spot sensitization mechanism is attempted.

This paper is organized as follows. In Sect.~\ref{sec2}, the governing equations and various scenarios considered in the simulations are stated. Section~\ref{sec3} describes the herein used numerical methodology in detail. Simulation results and analysis are presented in Sect.~\ref{sec4}. The findings based on the results, limitations of the current work, and future development of meso-resolved simulations to further study the SDT phenomena in NM-GMB mixtures are discussed in Sect.~\ref{sec5} and summarized in Sect.~\ref{sec6}. Numerical and statistical convergence studies are reported in Appendices~\ref{append1} and~\ref{append2}, respectively.

\section{Problem statement}
\label{sec2}

\subsection{Governing equations}
\label{sec2_1}
The dynamic system at hand is simulated based on the Euler equations that are augmented to treat a reactive, multi-phase flow with two immiscible materials (i.e., liquid NM and air).~\cite{Michael2016JCP} The use of this mathematical framework has previously been explored by Michael and Nikiforakis to simulate single cavity collapse in liquid NM.~\cite{Michael2014DS,Michael2018_I,Michael2018_II,Michael2019Book} The governing equations are formulated as follows,
\begin{equation}
\begin{split}
\frac{\partial z_1 \rho_1}{\partial t} + \nabla \cdot \left(z_1 \rho_1 \mathbf{u} \right) = 0 \\
\frac{\partial z_2 \rho_2}{\partial t} + \nabla \cdot \left(z_2 \rho_2 \mathbf{u} \right) = 0 \\
\frac{\partial}{\partial t}\left( \rho \mathbf{u} \right) + \nabla \cdot \left(\rho \mathbf{u} \otimes \mathbf{u} \right) + \nabla p = 0 \\
\frac{\partial}{\partial t}\left( \rho E \right) + \nabla \cdot \left[ \left(\rho E + p \right) \mathbf{u} \right] = -z_2 \rho_2 K Q\\
\frac{\partial z_1}{\partial t} + \mathbf{u} \cdot \nabla z_1 = 0 \\
\frac{\partial}{\partial t} \left(z_2 \rho_2 \lambda \right) + \nabla \cdot \left(z_2 \rho_2 \lambda \mathbf{u} \right) = z_2 \rho_2 K
\end{split}
\label{Eq1}
\end{equation}
The air within cavities is considered as phase~$1$ and the liquid NM is considered as phase~2, which are denoted as subscripts ``$1$'' and ``$2$'', respectively. The volume fractions of air and NM are represented by $z_1$ and $z_2$, respectively, where $z_1 + z_2 = 1$. The total density $\rho$ of the mixture can thus be calculated as $\rho = z_1 \rho_1 + z_2 \rho_2$. The total specific energy is defined as $E = \frac{1}{2}u^2 + e$, where $e$ is the specific internal energy of the mixture, i.e., $\rho e = z_1 \rho_1 e_1 + z_2 \rho_2 e_2$. The reaction progress variable $\lambda$ evolves from $1$ to $0$, representing the mass fraction of unreacted NM. The time rate of change of $\lambda$ is denoted as $K$. The specific energy release of NM is represented by $Q$. At material interfaces, these two immiscible materials are considered to be in mechanical equilibrium but not thermal equilibrium, i.e., $p_1=p_2=p$, $\mathbf{u}_1=\mathbf{u}_2=\mathbf{u}$, but $T_1$ and $T_2$ might not be equal.

{Since the thermal diffusivity of liquid NM $\alpha$ is on the order of $0.1~\mathrm{\mu m}^2/\mathrm{\mu s}$, the characteristic time scale of heat diffusion over a cavity diameter of $100~\mathrm{\mu m}$, i.e.,  $\tau_\mathrm{d} = {d_\mathrm{c}}^2/\alpha$, is on the order of nearly $1 \times 10^5 \mathrm{\mu s}$. The time scale of the SDT process is expected to be on the order of $\mathrm{\mu s}$ for the range of input shock pressure ($\sim 10~\mathrm{GPa}$) in this study. It is therefore justifiable in this model to neglect heat diffusion, which is too slow to significantly affect the hot-spot-triggered combustion. Due to the same reason, heat diffusion was also neglected by Menikoff\cite{Menikoff2011SWJ} in modeling hot spot formation in NM mixed with solid beads. The effect of viscous heating in this system consisting of liquid- and gas-phase materials is in general less significant than the viscoplastic heating effect due to intergranular friction in polycrystalline solid explosives. Also, viscous heating is speculated to significantly increase the hot spot temperature for low incident shock pressures.\cite{Field1992hot} In this study, since relatively high input shock pressures ($> 7~\mathrm{GPa}$) are considered, material viscosity is neglected in this model in order to carry out more focused parametric studies.

\subsection{Equations of state (EoS)}
\label{sec2_2}
In this work, the unreacted NM and its products are described by the same Cochran-Chan EoS~\cite{Cochran1979,Saurel2009JCP}, which is expressed as follows,
\begin{equation}
p(\rho_2, e_2) = p_\mathrm{ref,2}(\rho_2) + \rho_2 \Gamma_{0,2} \left[e_2-e_\mathrm{ref,2}(\rho_2) \right]
\label{Eq2}
\end{equation}
where the reference pressure $p_\mathrm{ref,2}$ is given by
\begin{equation}
p_\mathrm{ref,2}(\rho_2) = \mathcal{A} \left(\frac{\rho_{0,2}}{\rho_2}\right)^{-\mathcal{C}} -\mathcal{B} \left(\frac{\rho_{0,2}}{\rho_2}\right)^{-\mathcal{D}}
\label{Eq3}
\end{equation}
the reference energy $e_\mathrm{ref,2}$ is given by
\begin{equation}
\begin{split}
e_\mathrm{ref,2}(\rho_2) = & \frac{-\mathcal{A}}{\rho_{0,2}(1-\mathcal{C})} \left[ \left(\frac{\rho_{0,2}}{\rho_2}\right)^{1-\mathcal{C}} -1 \right] \\
+ & \frac{\mathcal{B}}{\rho_{0,2}(1-\mathcal{D})} \left[ \left(\frac{\rho_{0,2}}{\rho_2}\right)^{1-\mathcal{D}} -1 \right]
\end{split}
\label{Eq4}
\end{equation}
and $\Gamma_{0,2}$ is the Gr\"{u}neisen coefficient corresponding to the initial state of NM at $\rho_2 = \rho_{0,2}$. This EoS has been broadly used in the literature.~\cite{Saurel2009JCP,Shukla2010,Genetier2014,Michael2018_I,Michael2018_II,Michael2019Book} Air inside the cavities is governed by the ideal gas law, i.e., $e_1 = p/(\gamma_1 - 1)\rho_1$, {where $\gamma_1$ is the ratio of specific heat capacities of air}. The EoS of the ideal gas law can be equivalently expressed in the Mie-Gr\"{u}neisen form with $p_\mathrm{ref,1}=0$ and $e_\mathrm{ref,1}=0$. The values of $\gamma_1$ and the constant-volume specific heat capacity $c_\mathrm{v,1}$ of air are given in Table~\ref{Tab2}. 

{
The temperature of each material needs to be calculated separately according to its equation of state. A general expression for calculating the temperature of each material based on the first-order Taylor expansion from the reference curve should be as follows
\begin{equation}
    T_i - T_{\mathrm{ref},i}(\rho_i) = \frac{p - p_{\mathrm{ref},i}(\rho_i)}{\rho_i \Gamma_{i}(\rho_i,T_i) c_{\mathrm{v},i}(\rho_i,T_i)}  \;\;\;\;\;\;\mathrm{for} \;\;i=1,2
\end{equation}
where the constant-volume specific heat capacity $c_{\mathrm{v},i}$ and  Gr\"{u}neisen coefficient $\Gamma_{i}$ are functions of both density and temperature.\cite{Winey2000} The Cochran-Chan EoS used for liquid NM is based on the reference curves of isotherms, thus, $T_{\mathrm{ref},2}(\rho_2) = T_{\mathrm{ref},2}$.\cite{Cochran1979}} {The parameters, $\mathcal{A}$, $\mathcal{B}$, $\mathcal{C}$, and $\mathcal{D}$, in the functions for reference pressure and energy are calibrated using the approach proposed by Cochran and Chan~\cite{Cochran1979} to reproduce the Hugoniot relation between the experimental data of shock and particle velocities. The parameters $\mathcal{A}$ and $\mathcal{B}$ have dimensions of pressure, while $\mathcal{C}$ and $\mathcal{D}$ are dimensionless.} {In this study, the parameter values for this EoS (summarized in Table~\ref{Tab1}) are adopted from the literature~\cite{Michael2018_I,Michael2018_II} imposing a temperature $T_2 = 298~\mathrm{K}$ for liquid NM at its initial density $\rho_2 = \rho_{0,2}$, which is a physically accurate value for the initial temperature of NM. Thus, the Cochran-Chan isothermal reference curve is at a temperature $T_\mathrm{ref,2}=0~\mathrm{K}$. 

As described in this section, two major simplifications have been made in this model: (1) The Cochran-Chan EoS calibrated for liquid NM is also used to describe the reaction products; (2) the constant-volume specific heat capacities for both materials and the Gr\"{u}neisen coefficient for NM are considered to be constant values corresponding to the initial state of the materials, i.e., $c_{\mathrm{v},i}(\rho_i,T_i)=c_{\mathrm{v},i}$ and $\Gamma_{2}(\rho_2,T_2) = \Gamma_{0,2}$, respectively. Simplification (1) is made to avoid the iterative root-finding procedure to determine the volume fractions of reactant and products in the reaction zone, thus, ensuring the robustness and time-efficiency of the simulations to explicitly resolve a large number of heterogeneities. It is of importance to note that both of these simplifications impose a limitation on the quantitative accuracy of the model, which is further discussed in Sect.~\ref{sec5_5}.
}

\begin{table}
\begin{center}
\caption{Parameters for the equation of state of air}
\label{Tab2}
\begin{tabular}{| c | c |}
\hline
Parameter & Value (unit)\\
\hline
$\gamma_1$ & $1.4$ (-)\\
$c_\mathrm{v,1}$ & $718$ ($\mathrm{J}$ $\mathrm{kg}^{-1}\mathrm{m}^{-3}$)\\
\hline
\end{tabular}
\end{center} 
\end{table}

\begin{table}
\begin{center}
\caption{Parameters for the equation of state of liquid NM}
\label{Tab1}
\begin{tabular}{| c | c |}
\hline
Parameter & Value (unit)\\
\hline
$\Gamma_{0,2}$ & $1.19$ (-)\\
$\mathcal{A}$ & $0.819$ (GPa)\\
$\mathcal{B}$ & $1.51$ (GPa)\\
$\mathcal{C}$ & $4.53$ (-)\\
$\mathcal{D}$ & $1.42$ (-)\\
$\rho_{0,2}$ & $1134$ ($\mathrm{kg}$ $\mathrm{m}^{-3}$)\\
$c_\mathrm{v,2}$ & $1714$ ($\mathrm{J}$ $\mathrm{kg}^{-1}\mathrm{m}^{-3}$)\\
\hline
\end{tabular}
\end{center} 
\end{table}

\subsection{Reaction rate model}
\label{sec2_3}
The reaction rate $K$ of liquid NM is governed by single-step Arrhenius kinetics as follows,
\begin{equation}
K = \frac{\partial \lambda}{\partial t} = -\lambda C \mathrm{exp} (-T_\mathrm{a}/T_2)
\label{Eq6}
\end{equation}
{where $T_\mathrm{a}$ is the activation temperature and $C$ is the pre-exponential factor. The value of $Q$ was calculated by Michael and Nikiforakis\cite{Michael2018_II} following the approach proposed by Arienti \textit{et al}.~\cite{Arienti2004}:  Via varying the value of $Q$, the reactive Hugoniot curve can be shifted on the $p$-$v$ plane to match as closely as possible the CJ pressure of $12.5~\mathrm{GPa}$ reported by Dattelbaum \textit{et al.}~\cite{Dattelbaum2009APS}. The value $Q=4.46\times10^6 \mathrm{J}/\mathrm{kg}$ has thus been obtained. Based on the velocity interferometer data obtained for the range of shock pressure from $7.5$ to $9.5~\mathrm{GPa}$, Hardesty\cite{Hardesty1976} obtained $C=2.6 \times 10^9~\mathrm{s}^{-1}$ and $T_\mathrm{a}=11500~\mathrm{K}$ for a reaction rate in the first-order, single-step Arrhenius form. Fixing the value of $C$ as calibrated by Hardesty\cite{Hardesty1976}, $T_\mathrm{a}$ was re-calibrated by Michael and Nikiforkais\cite{Michael2018_II} using the obtained value of $Q$ and the same EoS as that used in this study to match the \textit{in-situ} gauge data of particle velocity reported by Sheffield \textit{et al.}~\cite{Sheffield2006} for neat NM shocked at $9.1~\mathrm{GPa}$. Thus, the value of $T_\mathrm{a} = 11350~\mathrm{K}$ has been obtained and used in this work. The values for $C$, $T_\mathrm{a}$, and $Q$ are summarized in Table~\ref{Tab3}.}

\begin{table}
\begin{center}
\caption{Parameters for the reaction law of liquid NM}
\label{Tab3}
\begin{tabular}{| c | c |}
\hline
Parameter & Value (unit)\\
\hline
$T_\mathrm{a}$ & $11350$ (K)\\
$C$ & $2.6 \times 10^9$ ($\mathrm{s}^{-1}$)\\
$Q$ & $4.46 \times 10^6$ ($\mathrm{J}$ $\mathrm{kg}^{-1}$)\\
\hline
\end{tabular}
\end{center} 
\end{table}

\subsection{Initial configuration of the problem}
\label{sec2_4}
Three different scenarios are considered in this study to examine the SDT behaviors: 1) Neat NM, 2) NM mixed with an array of regularly spaced circular cavities, and 3) NM mixed with randomly distributed cavities. The schematic illustration of these three scenarios is shown in Fig.~\ref{Fig1}. For all of the three scenarios, a rightward-moving incident shock wave is implemented via initially placing a shocked region near the left end of the domain with an inflow (transmissive) boundary condition applied upon the left boundary. Within this shocked region, the initial values of density, pressure, and particle velocity satisfy the Hugoniot relation for neat NM given the initial pre-shock conditions (see Sect.~5.1 in Michael's PhD thesis\cite{Michael2013}). The reaction progress variable is set to be $0$ in this shocked region so that this amount of initially shocked NM does not contribute to the global energy release of the mixture. The longitudinal length of the shocked region $L_\mathrm{s}$ (in the propagation direction of the incident shock) is chosen to be $0.5~\mathrm{mm}$ for all of the simulations reported in this paper. The overall length of the domain ahead of the incident shock front $L$ is chosen to be sufficiently long so that the complete SDT process can be captured for the corresponding incident shock strength.\\

\begin{figure*}
\centerline{\includegraphics[width=0.8\textwidth]{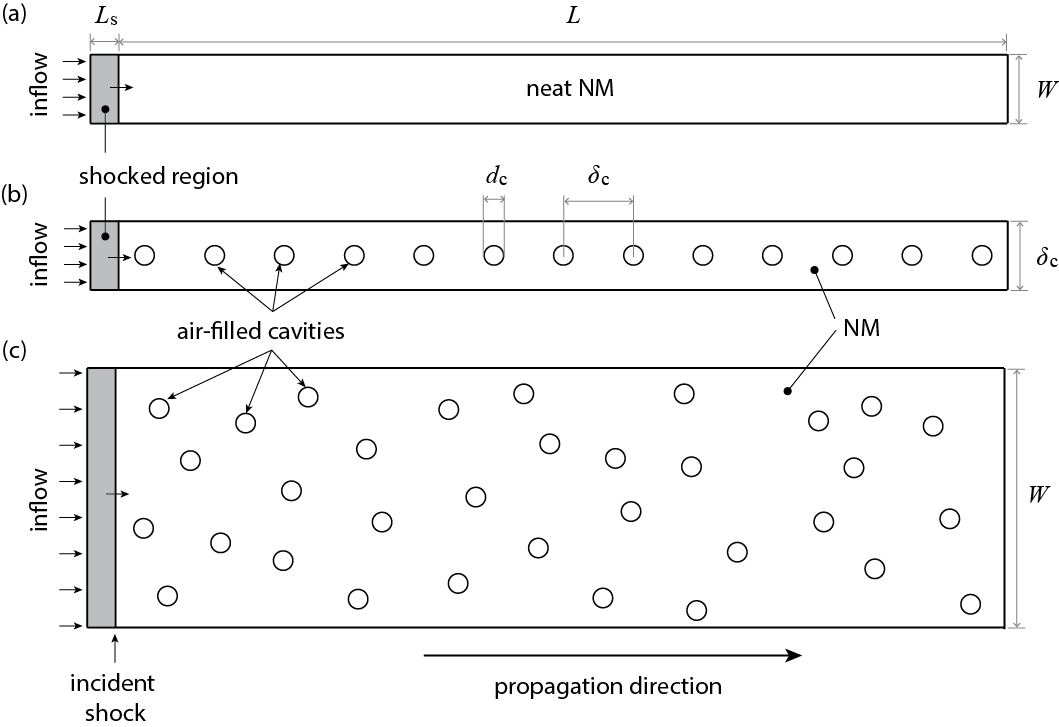}}
		\caption{Schematic illustration of the initial configuration of the computational domain for three difference scenarios: (a) Neat NM, (b) NM with an array of regularly spaced circular cavities, and (c) NM with randomly distributed cavities. Periodic boundary conditions are applied on the top and bottom boundaries of the domain.}
	\label{Fig1}
\end{figure*}

For the scenario with neat NM, since the resulting wave dynamics is expected to be one-dimensional, the transverse width of the domain $W$ is arbitrarily chosen to be $0.3~\mathrm{mm}$. For the case with regularly spaced cavities, as shown in Fig.~\ref{Fig1}(b), each cavity is of the same diameter $d_\mathrm{c}$ and spaced from its neighboring cavities by an equal distance $\delta_\mathrm{c}$. Specifying the overall porosity of the mixture $\phi$ (i.e., initial volume fraction of the air-filled cavities) and cavity diameter, the spacing between each two adjacent cavities can be calculated as follows,
\begin{equation}
\label{Eq6_1}
\delta_\mathrm{c} = \sqrt{\frac{\pi {d_\mathrm{c}}^2}{4\phi}}
\end{equation}
The transverse width of the domain for the scenario of regularly distributed cavities is set equal to $\delta_\mathrm{c}$ and periodic boundary conditions are applied on the top and bottom boundaries of the domain so that a two-dimensional array of equally spaced cavities can be simulated.

In the third scenario as shown in Fig.~\ref{Fig1}(c), the cavities are randomly distributed in a medium of NM. Knowing the total volume of the reactive mixture, i.e., $V_\mathrm{tot} = L\,W$, the total number of cavities in the domain can be calculated as
\begin{equation}
\label{Eq7}
N_\mathrm{tot} = \frac{4 \phi L\,W}{\pi {d_\mathrm{c}}^2}
\end{equation}
In each simulation, the cavities are of the same size. The initial pressure and density of the unshocked air and NM are $p=10^5~\mathrm{Pa}$, $\rho_1 = 1.2~\mathrm{kg}/\mathrm{m}^3$, and $\rho_2 = 1134~\mathrm{kg}/\mathrm{m}^3$, respectively, for all of the simulations in this study. The total mass of reactive NM in the system can thus be calculated as $M_{2,\mathrm{tot}} = \rho_2 (1-\phi) L\,W$.

\section{Numerical methodology}
\label{sec3}

The simulation code used to solve the two-dimensional, reactive Euler equations is based upon a uniform Cartesian grid. The MUSCL-Hancock scheme with the van Leer non-smooth slope limiter and a Harten-Lax-van Leer-contact (HLLC) approximate solver for the Riemann problem were used.~\cite{Toro2009} The Strang splitting method was adopted to treat separately the hydrodynamic process and the reactive process. This numerical scheme is thus of second-order accuracy in space and time. 

The simulation code was implemented in NVIDIA's CUDA programming language to take the advantage of Graphic-Processing-Unit- (GPU-) enabled parallel computing. Each simulation was performed with a $16$-GB NVIDIA Tesla P100 GPU. The use of GPU-accelerated computing platforms has been explored in several studies on gaseous detonations.~\cite{Morgan2013,KiyandaNg2015,Mi2017PRF,Mi2018SWJ} The current code is a further development of the gaseous-detonation code to simulate detonations in a multiphase energetic system. This code is of a hybrid nature, consisting of parts that are executed on both CPU (Central Processing Unit) and GPU. Under a Godunov-type numerical scheme, CUDA kernel functions were implemented to parallelize the most compute-intensive parts within the solver for the reactive Euler equations, including computation of fluxes, reactive source term integration, and the update of conserved variables. In order to minimize dataflow between the global memory of a GPU and its streaming multiprocessors, the whole procedure of flux computation was implemented in such a way using the on-chip, shared memory of a GPU. Data transfer between host memory and GPU global memory is only required for domain initialization and data output at prescribed times. The code for post-simulation analysis was implemented in a serial fashion and executed on the host CPU.

A diffuse-interface approach was used in this work where two immiscible materials are separated by a thin numerical diffusion layer. Comparing to interface-tracking methodologies wherein a sharp material interface is resolved, a diffuse-interface approach demands a significantly lower computational cost. A sufficient grid resolution is however required to ensure a sufficiently thin diffusion layer at material interfaces. The validation of using this approach to simulate a single or a small number of air-filled cavities or solid particles under shock loading in liquid NM has been demonstrated in several previous studies.\cite{Michael2013,Michael2016JCP,Michael2018_I,Michael2018_II,Michaell2019Book} In this work, since a large number of  cavities are present in the simulations, the total area covered by numerical diffusion zones was greater than that in previous studies. In order to avoid any numerical issues that can probably arise from entangled diffuse interfaces from neighboring collapsed cavities, a minimum spacing distance of $2.5 d_\mathrm{c}$ was imposed to the random distribution of cavities. Further validation tests have been performed, and the computational grid-size independence of the key results has been verified and reported in Appendix~~\ref{append1}.

\section{Results and analysis}
\label{sec4}

For the cases with air-filled cavities considered in this study, the overall porosity and cavity diameter were selected to be $\phi=8\%$ and $d_\mathrm{c}=100~\mu\mathrm{m}$. {This selected value of cavity diameter is very close to the range of mean diameter of GMBs used to sensitize gelled NM as reported in a number of experimental studies.\cite{Gois1996,Gois2002,Dattelbaum2010,Dattelbaum2010Role,Higgins2013APS,Higgins2018APS,Loiseau2018SWJ}} The detailed comparison between these selected parameter values to realistic conditions of experimental studies on SDT in GMB-sensitized NM will be discussed in Sect.~\ref{sec5_5}. Sample results showing the evolution of wave structure for the scenarios with neat NM, regularly spaced cavities, and randomly distributed cavities are shown in Sect.~\ref{sec4_1} as contour plots of NM density, i.e., $z_2 \rho_2$. To illustrate the difference in SDT behaviors in neat and heterogeneous NM mixtures, in Sect.~\ref{sec4_2} and \ref{sec4_3_new}, the corresponding spatio-temporal profiles of pressure and particle velocity, respectively, are compared. In Sect.~\ref{sec4_3}, the global reaction rate as a function of time is shown for each scenario. Based on the time history of the global reaction rate, a characteristic time scale of the resulting SDT process, or the so-called ``detonation overtake time'', can be measured. The simulation results of overtake time as a function of incident shock pressure are presented and compared to available experimental data of overtake time for GMB-sensitized, gelled NM in Sect.~\ref{sec4_4}. Statistical analysis on the simulation data has been performed and reported in Sect.~\ref{sec4_5} to further investigate the collective effect of hot spots. To obtain the results reported in this section, simulations were performed at a numerical resolution of $\mathrm{d}x=\mathrm{d}y=1~\mu\mathrm{m}$, ensuring $100$ computational points across the diameter of a cavity. With evidence provided in the Appendices~\ref{append1} and~\ref{append2}, it has been verified, respectively, that further refining the computational grid or increasing the domain size in the $y$-direction would not alter the key findings in this study. {It is worthwhile to remark that, as calculated by Michael and Nikiforkais\cite{Michael2018_II} (shown in Fig.~1 of their paper) using the same model for neat liquid NM, the steady Zel'dovich-von Neumann-D\"{o}ring (ZND) detonation profile has a reaction zone length of approximately $200~\mu\mathrm{m}$, which is twice the cavity diameter $d_\mathrm{c}=100~\mu\mathrm{m}$ selected in this paper. At the nominal grid resolution of $\mathrm{d}x=1~\mu\mathrm{m}$ used in this study, there are thus nearly $200$ computational cells per reaction zone length of the ZND solution.}

\subsection{Wave structure}
\label{sec4_1}

The evolution of the wave structure over the course of an SDT process in neat NM, mixture with an array of regularly spaced cavities, and a mixture with randomly distributed cavities is shown in Figs.~\ref{Fig2}, \ref{Fig3}, and \ref{Fig4}, respectively. In these figures, the color contours represent the spatial distribution of NM density ($\rho_2 z_2$). Each rectangular subfigure is a snapshot of the entire two-dimensional computational domain at a specific time. These subfigures are arranged from bottom to top in a chronological order. The incident shock pressure for the selected sample results is $7.81~\mathrm{GPa}$. The length of the domain is $24~\mathrm{mm}$.

\begin{figure*}
\centerline{\includegraphics[width=0.9\textwidth]{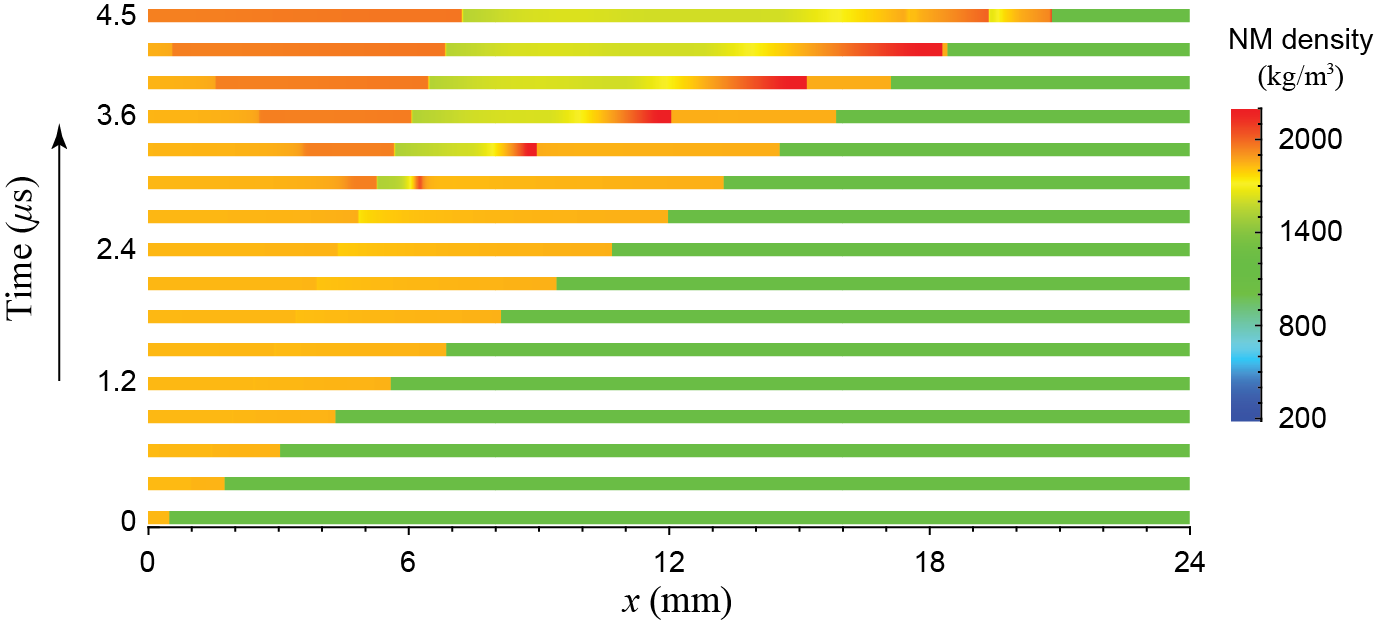}}
		\caption{Color contour plots of NM density ($z_2 \rho_2$) showing the evolution of wave structure over the SDT process in neat NM subjected to an incident shock pressure of $7.81~\mathrm{GPa}$. The subplots are arranged from bottom to top in a chronological order with a time interval of $0.3~\mu\mathrm{s}$.}
	\label{Fig2}
\end{figure*}

As shown in Fig.~\ref{Fig2}, the wave structure resulting from the case with neat NM is one-dimensional. The time interval between each two consecutive snapshots is $0.3~\mu\mathrm{s}$. From the beginning ($t=0$), the incident shock wave can be identified as a rightward-moving abrupt change in density. From $t=3.0~\mu\mathrm{s}$ onwards, a high-density region appears significantly downstream from the incident shock front, e.g., at $t=2.7~\mu\mathrm{s}$, the incident shock has reached approximately $x=13~\mathrm{mm}$ and this downstream high-density region is at $x=6~\mathrm{mm}$. This high-density front moves rightwards with a trailing gradient and catches up with the incident shock slightly after $t=4.2~\mu\mathrm{s}$.

\begin{figure*}
\centerline{\includegraphics[width=0.9\textwidth]{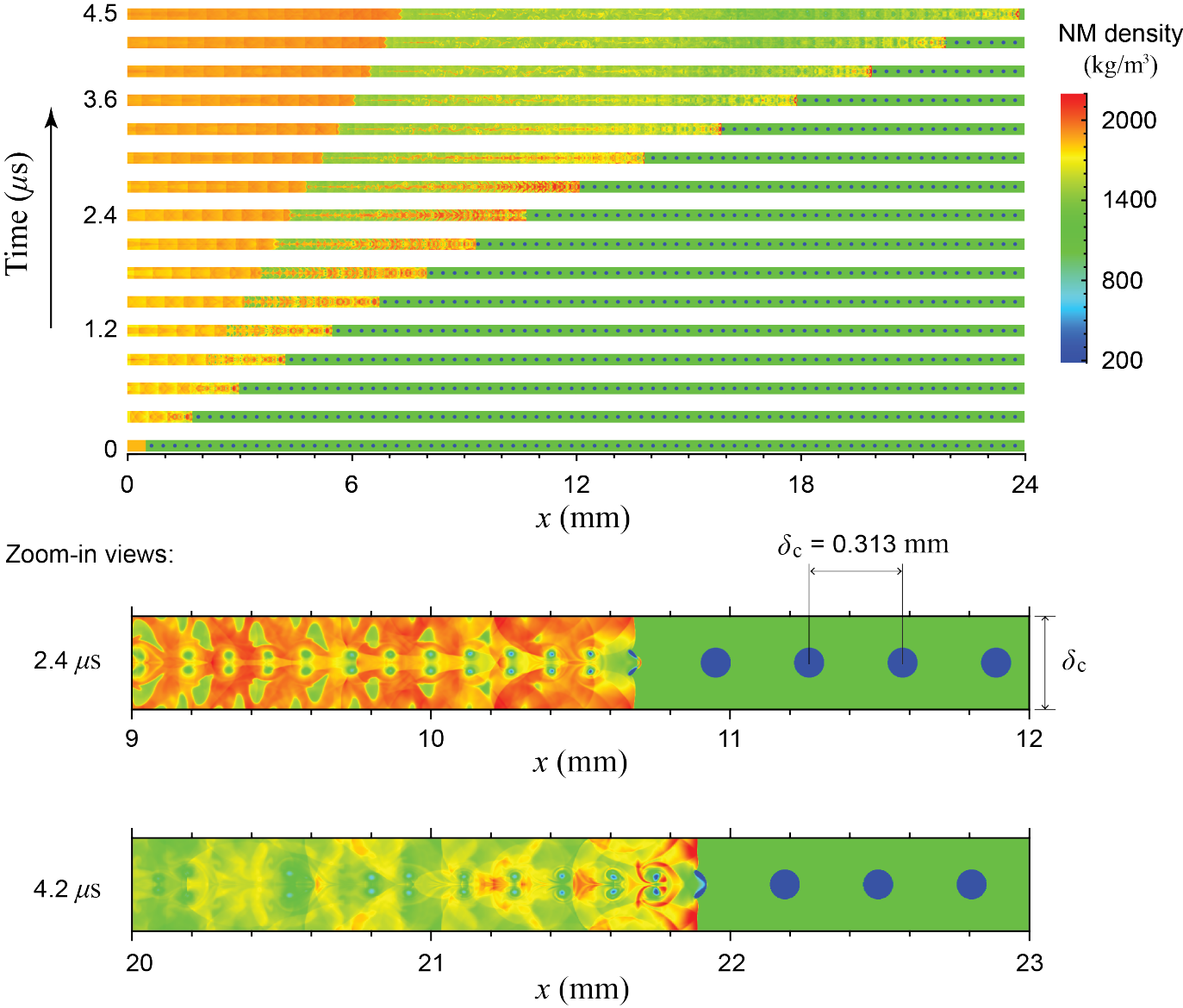}}
		\caption{Color contour plots of NM density ($z_2 \rho_2$) showing the evolution of wave structure over the SDT process in a mixture of NM and an array of regularly spaced cavities of $\phi=8\%$ and $d_\mathrm{c}=100~\mu\mathrm{m}$ subjected to an incident shock pressure of $7.81~\mathrm{GPa}$. The subplots are arranged from bottom to top in a chronological order with a time interval of $0.3~\mu\mathrm{s}$ with two zoom-in views showing the field of NM density in the vicinity of the leading shock front at $t=2.4~\mu\mathrm{s}$ and $4.2~\mu\mathrm{s}$.}
	\label{Fig3}
\end{figure*}

Snapshots showing the evolution of wave structure resulting from the scenario wherein a regular array of air-filled cavities embedded within a reactive matrix of NM are provided in Fig.~\ref{Fig3}. Zoom-in views of the flow field near the leading shock front at two different times, i.e., $t=2.4~\mu\mathrm{s}$ and $4.2~\mu\mathrm{s}$, are also shown in Fig.~\ref{Fig3}. For $\phi=8\%$ and $d_\mathrm{c}=100~\mu\mathrm{m}$, the spacing between each two neighboring cavities is $\delta_\mathrm{c}=0.313~\mathrm{mm}$. The resulting wave structure is two-dimensional and symmetric about the centerline along which the centers of the cavities are located. From $t=0$ to $t=2.7~\mu\mathrm{s}$, a region with non-uniformly distributed high-density loci (with $z_2 \rho_2 > 2000~\mathrm{kg}/\mathrm{m}^3$) develops immediately behind the incident shock and spans over a significantly long distance downstream. The zoom-in view at $t=2.4~\mu\mathrm{s}$ reveals more details of this high-density region coupled to the leading shock front. From $t=3~\mu\mathrm{s}$ onwards, high-density loci can be found only in the vicinity of the incident shock interacting with a collapsing cavity (as shown in the zoom-in view at $t=4.2~\mu\mathrm{s}$). The high-density wave front that develops significantly downstream from the incident shock in the case with neat NM is not observed in the case where a regular array of cavities is present.

\begin{figure*}
\centerline{\includegraphics[width=0.9\textwidth]{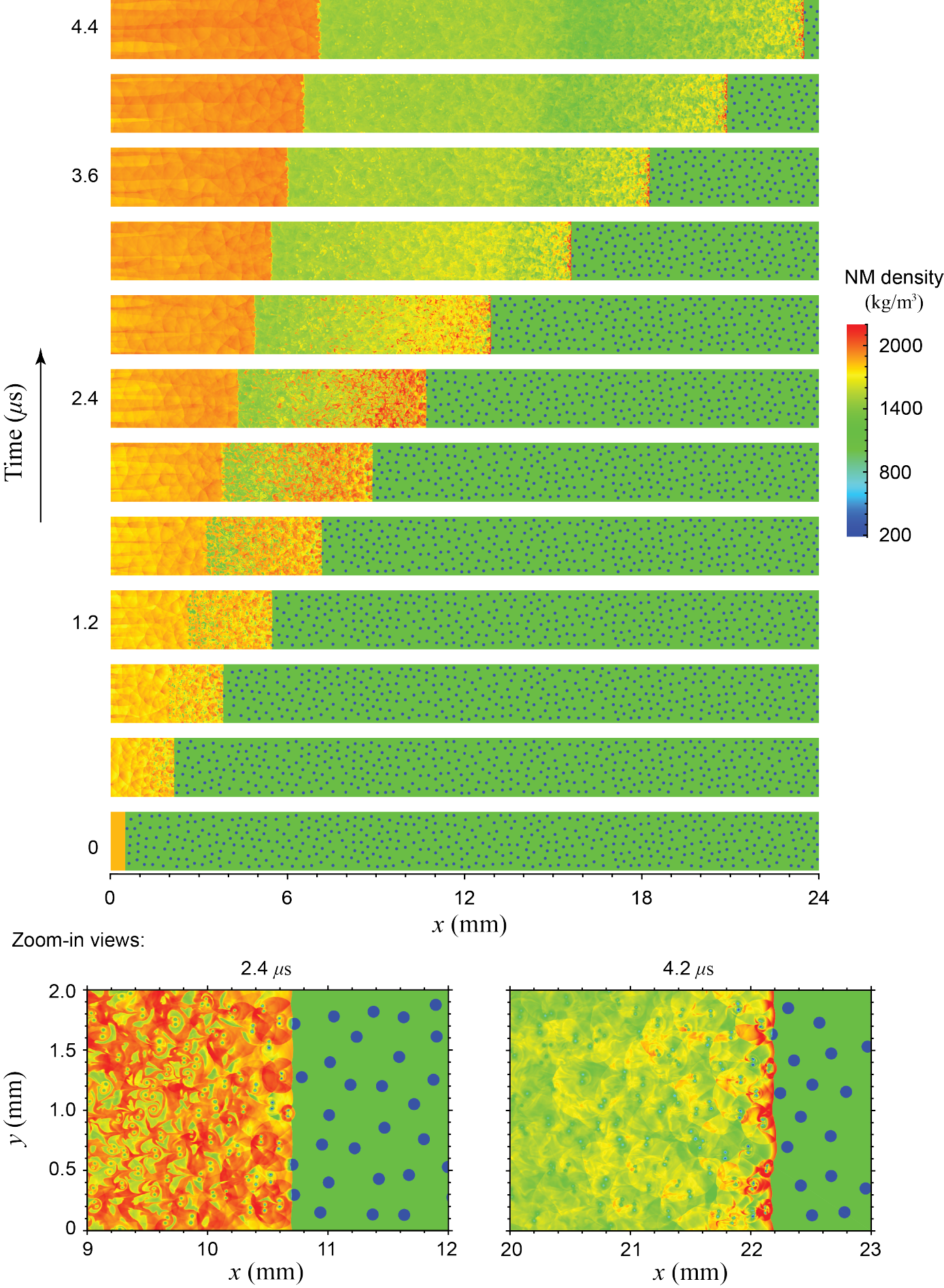}}
		\caption{Color contour plots of NM density ($z_2 \rho_2$) showing the evolution of wave structure over the SDT process in a mixture of NM and randomly distributed cavities of $\phi=8\%$ and $d_\mathrm{c}=100~\mu\mathrm{m}$ subjected to an incident shock pressure of $7.81~\mathrm{GPa}$. The subplots are arranged from bottom to top in a chronological order with a time interval of $0.4~\mu\mathrm{s}$ with two zoom-in views showing the field of NM density in the vicinity of the leading shock front at $t=2.4~\mu\mathrm{s}$ and $4.2~\mu\mathrm{s}$.}
	\label{Fig4}
\end{figure*}

The evolution of the wave structure for the scenario with a spatially random distribution of cavities is shown in Fig.~\ref{Fig4}. The overall wave structure is qualitatively similar to that observed in the case with regularly distributed cavities: A region with high-density loci is closely attached to the incident shock from $t=0$ to approximately $t=2.7~\mu\mathrm{s}$; afterward, high-density loci are only present at the leading shock front as shown in the zoom-in view at $t=4.2~\mu\mathrm{s}$. Due to the intrinsic random distribution of the cavities, the resulting wave structure is highly irregular compared to that for the case with a regular distribution of cavities. The feature observed in homogeneous NM---a downstream developed high-density wave front catching up with the leading shock---is not observed here. Further analysis will be performed to elucidate the different behaviors resulting from neat and cavity-laden NM mixtures.

\subsection{$x$-$t$ diagrams}
\label{sec4_2}

To further analyze the simulation results of the SDT process in neat NM, color contours of pressure and particle velocity in the $x$-direction for the case with an incident shock of $7.81~\mathrm{GPa}$ are plotted in an $x$-$t$ diagram as shown in Figs.~\ref{Fig5}(a) and \ref{Fig5_new}(a), respectively. The trajectory of the incident shock wave can be identified as a sharp interface separating a low-pressure (black) region on the right and a relatively high-pressure (dull red) region on the left starting from the bottom of this diagram at $t=1~\mu\mathrm{s}$. The trajectory of the interface between the inflow inert material and the explosive is denoted by white dashed line in Figs.~\ref{Fig5}(a). In the shocked region behind the incident shock wave, the trajectory of another shock wave can be identified in Fig.~\ref{Fig5}(a) as a sharp interface with a very high-pressure (bright yellow) region on its left originating at roughly a locus of $x=5~\mathrm{mm}$ and $t=2.8~\mu\mathrm{s}$ on the trajectory of the inflow material interface. The pressure behind (on the left of) this second shock evidently increases over time from $t=2.8~\mu\mathrm{s}$ to $3.4~\mu\mathrm{s}$. A fan-shaped region of decreasing pressure in the negative $x$-direction is trailed by this second shock wave. The trajectories of the incident shock wave and the second shock wave intersect at $t=4.22~\mu\mathrm{s}$, suggesting that the second shock wave overtakes the incident shock wave at this time. Thus, $t=4.22~\mu\mathrm{s}$ can be considered as the simulation result of the overtake time of the SDT process in neat NM for an incident shock pressure of $7.81~\mathrm{GPa}$. After the overtake, a single shock front trailing a decreasing pressure gradient can be found near the top-right corner of this diagram. 

\begin{figure}
\centerline{\includegraphics[width=0.47\textwidth]{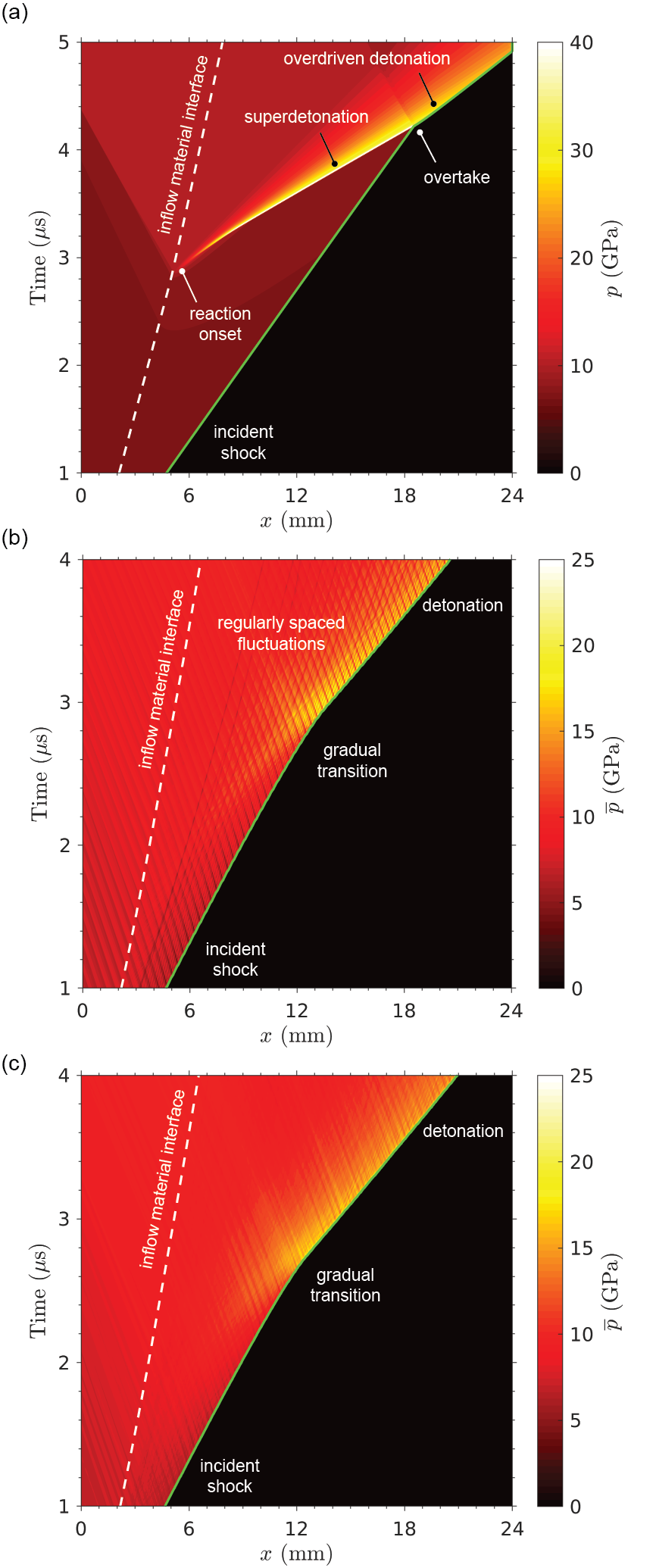}}
		\caption{Color contour plots of (a) pressure $p$ for the case of neat NM (case shown in Fig.~\ref{Fig2}) and spatially averaged pressure $\bar{p}$ for the case of NM mixed with (b) regularly and (c) randomly distributed cavities of $\phi=8\%$ and $d_\mathrm{c}=100~\mu\mathrm{m}$ (cases shown in Figs.~\ref{Fig3} and \ref{Fig4}, respectively) subjected to an incident shock pressure of $7.81~\mathrm{GPa}$ in $x$-$t$ diagrams.}
	\label{Fig5}
\end{figure}

As shown in Figs.~\ref{Fig3} and \ref{Fig4}, the wave dynamics resulting from the cases with cavities are two-dimensional. In order to examine the overall SDT process in an $x$-$t$ diagram, at a specific time, a one-dimensional profile (in the $x$-direction) of averaged flow and thermodynamic properties can be obtained via performing spatial averaging over the transverse direction (i.e., the $y$-direction) of the domain on the simulation data of the two-dimensional flow field. At time $t$, two-dimensional fields of pressure $p(x,y,t)$, density $\rho(x,y,t)$, and particle velocity in the $x$-direction $u(x,y,t)$ have been obtained from simulations. The one-dimensional profiles of average pressure $\bar{p}(x,t)$ and average density $\bar{\rho}(x,t)$ can be calculated as follows, respectively,
\begin{equation}
\bar{p}(x,t) = \frac{1}{W} \int_{0}^{W} p(x,y,t) \mathrm{d}y
\end{equation}
and
\begin{equation}
\bar{\rho}(x,t) = \frac{1}{W} \int_{0}^{W} \rho(x,y,t) \mathrm{d}y
\end{equation}
where $W$ is the width of the domain in the $y$-direction. With $\bar{\rho}(x,t)$ obtained, the density-weighted average profile of particle velocity $\bar{u}(x,t)$ be calculated as follows,
\begin{equation}
\bar{u}(x,t) = \frac{1}{W \bar{\rho}(x,t)} \int_{0}^{W} \rho(x,y,t) u(x,y,t) \mathrm{d}y
\end{equation}
The color contours of $\bar{p}$ for the cases with regularly and randomly distributed cavities subjected to an incident shock of $7.81~\mathrm{GPa}$ are thus plotted in an $x$-$t$ diagram as shown in Fig.~\ref{Fig5}(b) and (c), respectively. The results of $\bar{u}$ for the cavity-laden cases are used to obtain the time histories of particle velocity along particle paths (i.e., at Lagrangian positions) in Sect.~\ref{sec4_3_new}.

Comparing the plots shown in Fig.~\ref{Fig5}(b) and (c), the overall SDT behaviors resulting from both heterogeneous mixtures with regularly and randomly distributed cavities are qualitatively similar: Only one leading shock trajectory, i.e., a sharp interface separating a low-pressure region on the right and a high-pressure region on the left, can be identified. The slope of this shock trajectory gradually changes over the time span of these diagrams. A smooth, but rather evident, turning point in the shock trajectory can be found roughly from $t=2.8~\mu\mathrm{s}$ to $3.2~\mu\mathrm{s}$ for the case with regularly spaced cavities, and $t=2.4~\mu\mathrm{s}$ to $3~\mu\mathrm{s}$ for the case with randomly spaced cavities. The pressure behind this shock wave gradually changes. No second shock wave with a region of pressure significantly greater than $20~\mathrm{GPa}$  can be seen on the $x$-$t$ plot for the cavity-laden cases. For the case with a regular distribution of cavities, as shown in Fig.~\ref{Fig5}(b), some regularly spaced fluctuations due to shock-cavity interactions can be observed in the contour plot of average pressure. 

\subsection{Time history of particle velocity}
\label{sec4_3_new}
{
The time histories of particle velocity along a series of particle paths spanning through the domain in the $x$-direction are obtained from the $x$-$t$ data of particle velocity $u$ (or $\bar{u}$ for the cavity-laden cases). Since the overall SDT behaviors resulting from the cases with regularly and randomly distributed cavities are similar, only the results of particle velocity histories for neat NM and a random distribution of cavities are reported in this section. The contour plots of $u(x,t)$ and $\bar{u}(x,t)$ resulting from the cases with neat NM and a random distribution of cavities of $\phi=8\%$ and $d_\mathrm{c}=100~\mu\mathrm{m}$ subjected to an incident shock pressure of $7.81~\mathrm{GPa}$ are shown in Fig.~\ref{Fig5_new}(a) and (b), respectively. The red curves shown on these $x$-$t$ diagrams indicate the particle paths that are numerically tracked using forward Euler methods.

\begin{figure}
\centerline{\includegraphics[width=0.47\textwidth]{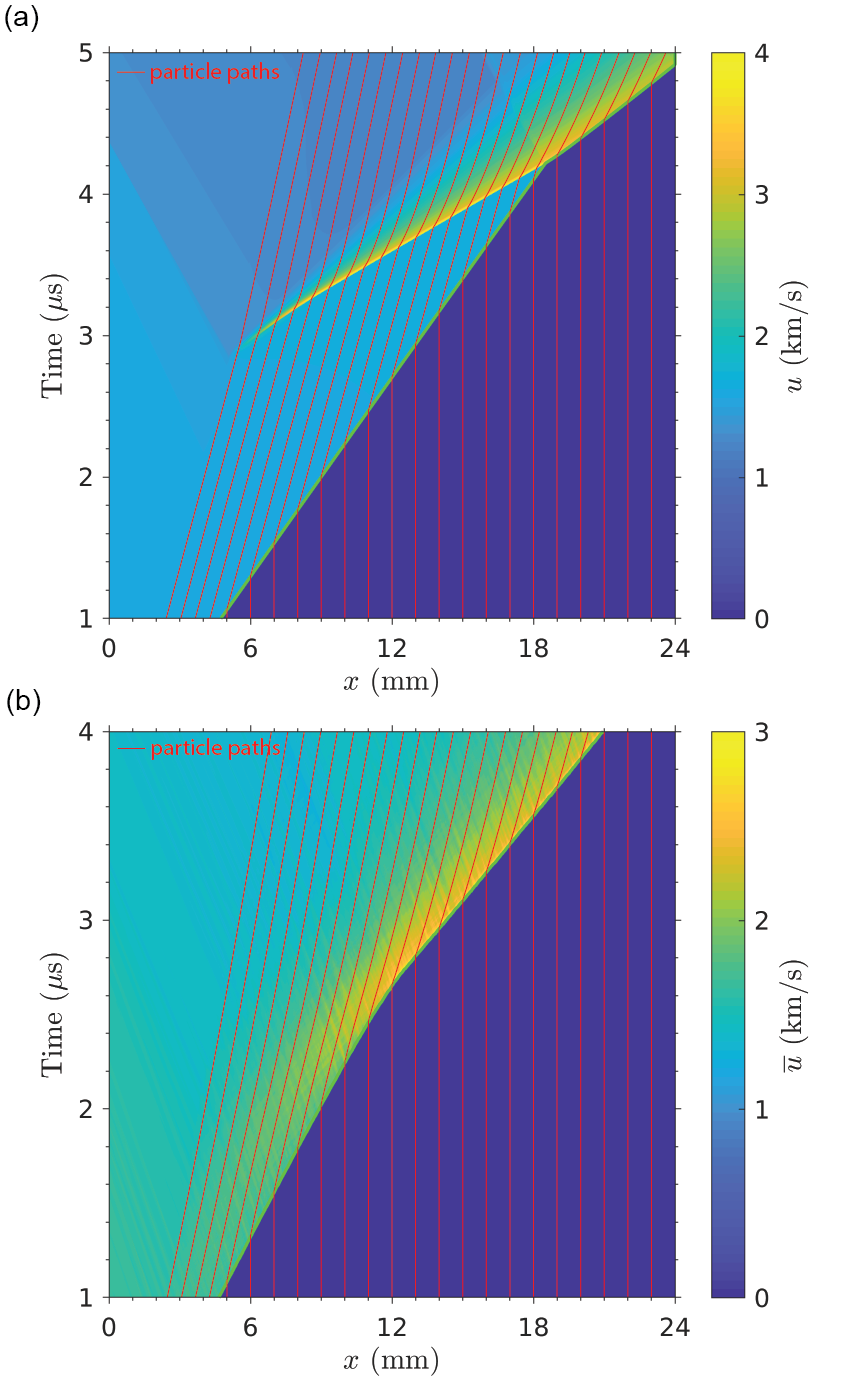}}
		\caption{Color contour plots of (a) flow velocity in the $x$-direction $u$ for the case of neat NM (case shown in Fig.~\ref{Fig2}) and (b) density-weighted, spatially averaged flow velocity $\bar{u}$ for the case of NM mixed with randomly distributed cavities of $\phi=8\%$ and $d_\mathrm{c}=100~\mu\mathrm{m}$ (case shown in Fig.~\ref{Fig4}) subjected to an incident shock pressure of $7.81~\mathrm{GPa}$ in $x$-$t$ diagrams.}
	\label{Fig5_new}
\end{figure}

The time histories of $u$ for the case with neat NM along a number of selected particle paths are shown in Fig~\ref{Fig_u_neat}. The location of each curve on the axis of $x_0$ indicates the initial position of the corresponding particle path. Subfigures (a) and (b) of Fig~\ref{Fig_u_neat} show the histories along the particle paths that are initially in the first (from $x=0$ to $12~\mathrm{mm}$) and second (from $x=12~\mathrm{mm}$ to $24~\mathrm{mm}$) halves of the domain, respectively. In each time history of $u$ shown in Fig.~\ref{Fig_u_neat}(a), there is an abrupt increase in $u$ at an early time followed by a period of constant velocity. At the end of this plateau, a second peak in $u$ occurs. The value of this second peak increases, and the time gap between the first increase and the second peak decreases from one particle path to the next. As shown in Fig~\ref{Fig_u_neat}(b), along the particle path initially located at $x=20~\mathrm{mm}$, the second peak in $u$ catches up to the first increase and merges into one abrupt increase. 

For the case with randomly distributed cavities, as shown in Fig.~\ref{Fig_u_random}, only one abrupt increase in $\bar{u}$ can be identified in the time history of $u$ along each particle path. This abrupt increase is followed by a gradual increase and decrease in particle velocity exhibiting a hump-shaped behavior. The peak value of this hump increases and its location becomes closer to the abrupt increase from one particle path to the next. The time history eventually evolves to a shape consisting of an abrupt increase followed by a gradually decreasing tail. As revealed by the contour plots in $x$-$t$ diagrams and the time histories of particle velocity, the SDT resulting from a cavity-laden NM mixture is a gradual process. It is rather inconvenient to measure a unique time scale to characterize this smooth transition process in heterogeneous NM, which is comparable to the overtake time for an SDT process in neat NM. Further analysis is required to obtain this characteristic time scale. 

\begin{figure*}
\centerline{\includegraphics[width=0.9\textwidth]{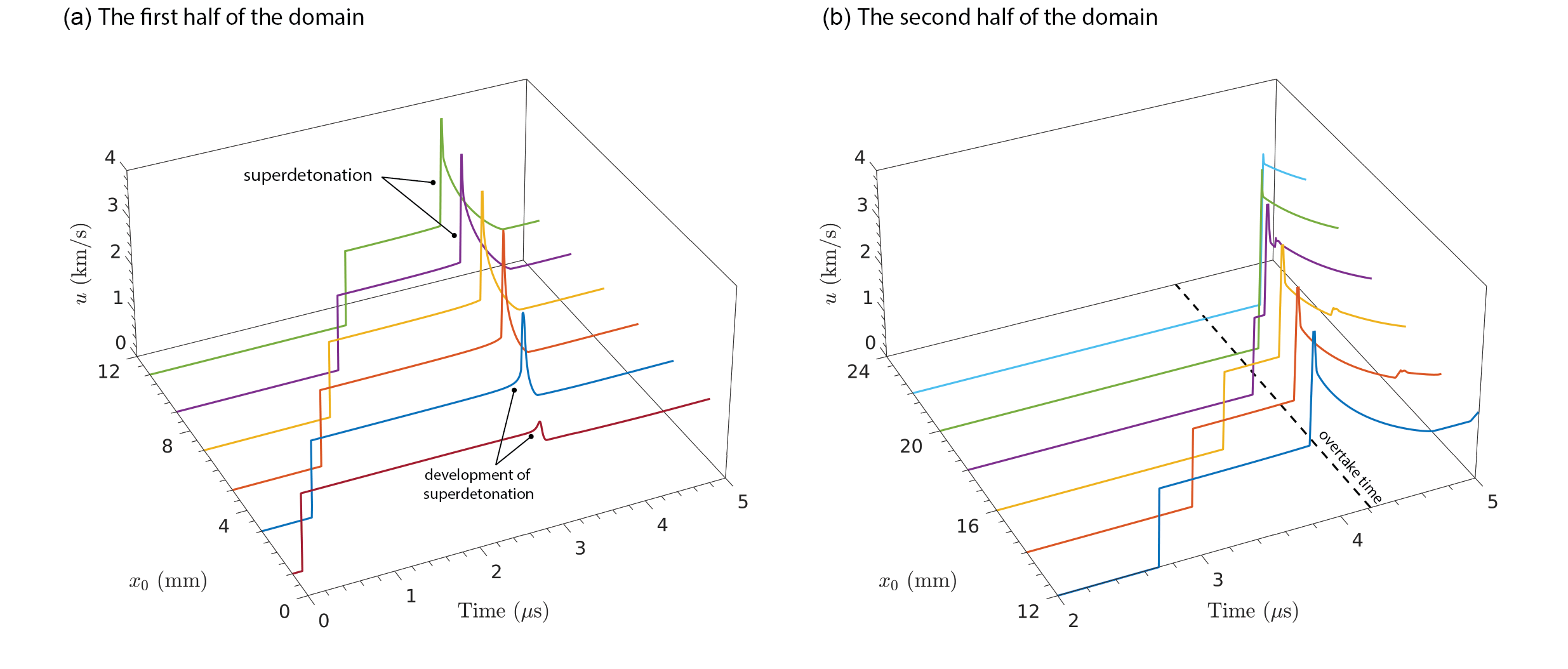}}
		\caption{The time histories of particle velocity $u$ along a series of particle paths (shown as the red curves in Fig.~\ref{Fig5_new}(a)) across the domain in the $x$-direction for the case with neat NM subjected to an incident shock pressure of $7.81~\mathrm{GPa}$ (i.e., the case shown in Fig.~\ref{Fig2}). The location of each curve on the axis of $x_0$ indicates the initial position of the corresponding particle path. Subfigures (a) and (b) show the histories along the particle paths that are initially in the first ($x=0$ to $12~\mathrm{mm}$) and second ($x=12~\mathrm{mm}$ to $24~\mathrm{mm}$) halves of the domain, respectively.}
	\label{Fig_u_neat}
\end{figure*}

\begin{figure}
\centerline{\includegraphics[width=0.47\textwidth]{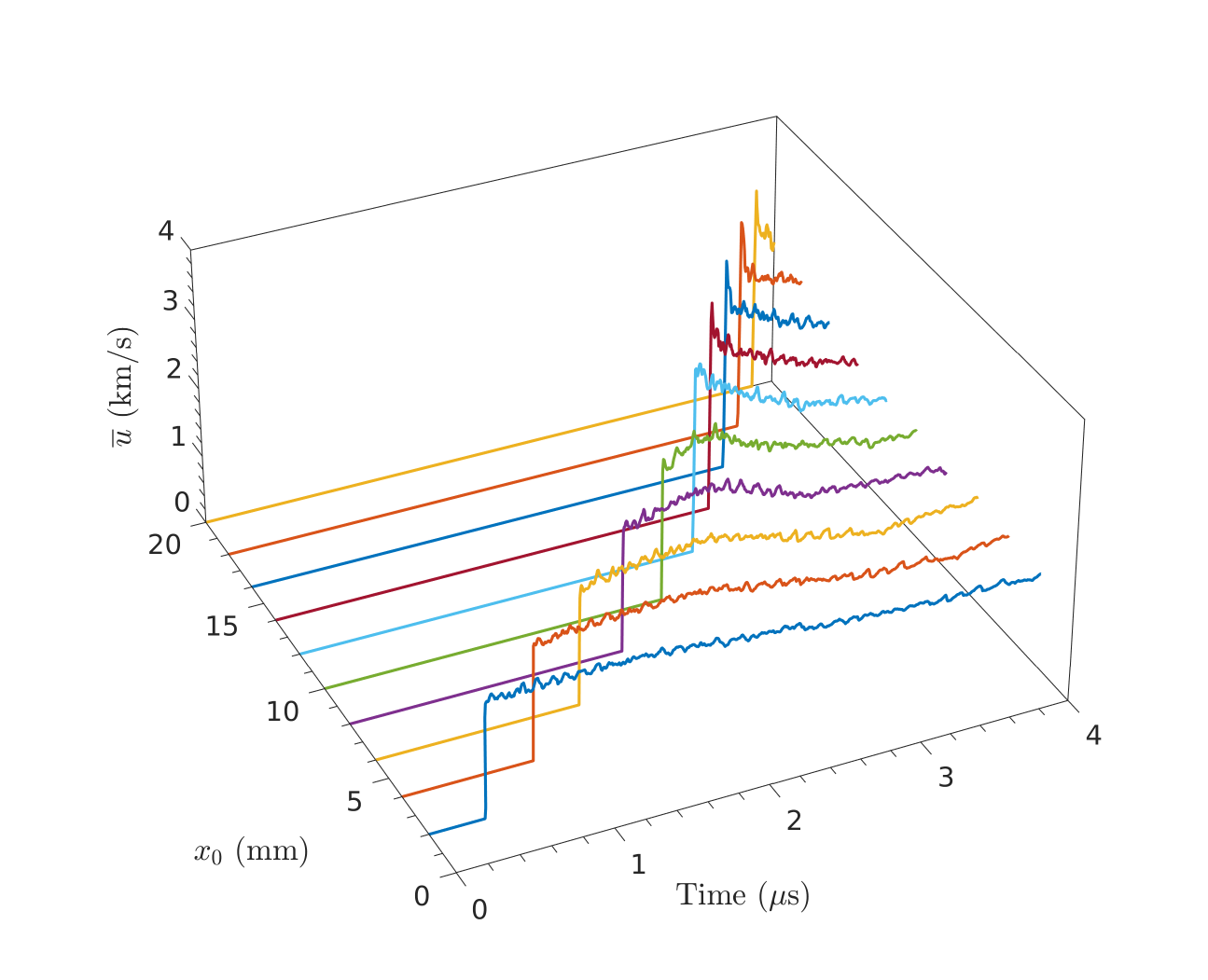}}
		\caption{The time histories of average particle velocity $\bar{u}$ along a series of particle paths (shown as the red curves in Fig.~\ref{Fig5_new}(b)) across the domain in the $x$-direction for the case of NM mixed with randomly distributed cavities of $\phi=8\%$ and $d_\mathrm{c}=100~\mu\mathrm{m}$ subjected to an incident shock pressure of $7.81~\mathrm{GPa}$ (i.e., the case shown in Fig.~\ref{Fig4}). The location of each curve on the axis of $x_0$ indicates the initial position of the corresponding particle path.}
	\label{Fig_u_random}
\end{figure}
}
\subsection{Global reaction rate}
\label{sec4_3}
Via analyzing the overall reaction rate of the mixture, this characteristic time, or the overtake time, for an SDT process in cavity-laden heterogeneous NM mixtures can be well defined, measured, and compared to that for the cases with neat NM. At each time step, the total mass of reactive NM remaining in the entire domain can be calculated via the following integration,
\begin{equation}
\label{Eq10}
M_{2,\mathrm{r}}(t) = \int_{0}^{W} \int_{0}^{L} \rho_2 z_2 \lambda \mathrm{d}x\mathrm{d}y
\end{equation}
The consumption rate of reactive NM in the entire domain can be obtained by differentiating $M_{2,\mathrm{r}}(t)$ over time, i.e., $\mathrm{d} M_{2,\mathrm{r}} / \mathrm{d} t$. For the simulations with the same incident shock strength but different scenarios of cavity distribution, the domain length $L$ is maintained the same, but domain width $W$ is different. Therefore, in order to compare the overall reaction progress resulting from cases with different domain widths on the same basis, the corresponding $\mathrm{d} M_{2,\mathrm{r}} / \mathrm{d} t$ needs to be normalized with respect to initial total mass of reactive NM. The normalization of the global reaction rate $\left | \bar{\lambda}_t \right|$ can be performed as follows,
\begin{equation}
\label{Eq11}
\left | \bar{\lambda}_t \right| = \left | \frac{\mathrm{d} M_{2,\mathrm{r}}}{\mathrm{d} t} / M_{2,\mathrm{tot}} \right |
\end{equation}
Since the consumption rate of reactive NM is negative, the absolute value of $\bar{\lambda}_t$ is taken so that the corresponding results can be plotted as positive values.

\begin{figure}
\centerline{\includegraphics[width=0.43\textwidth]{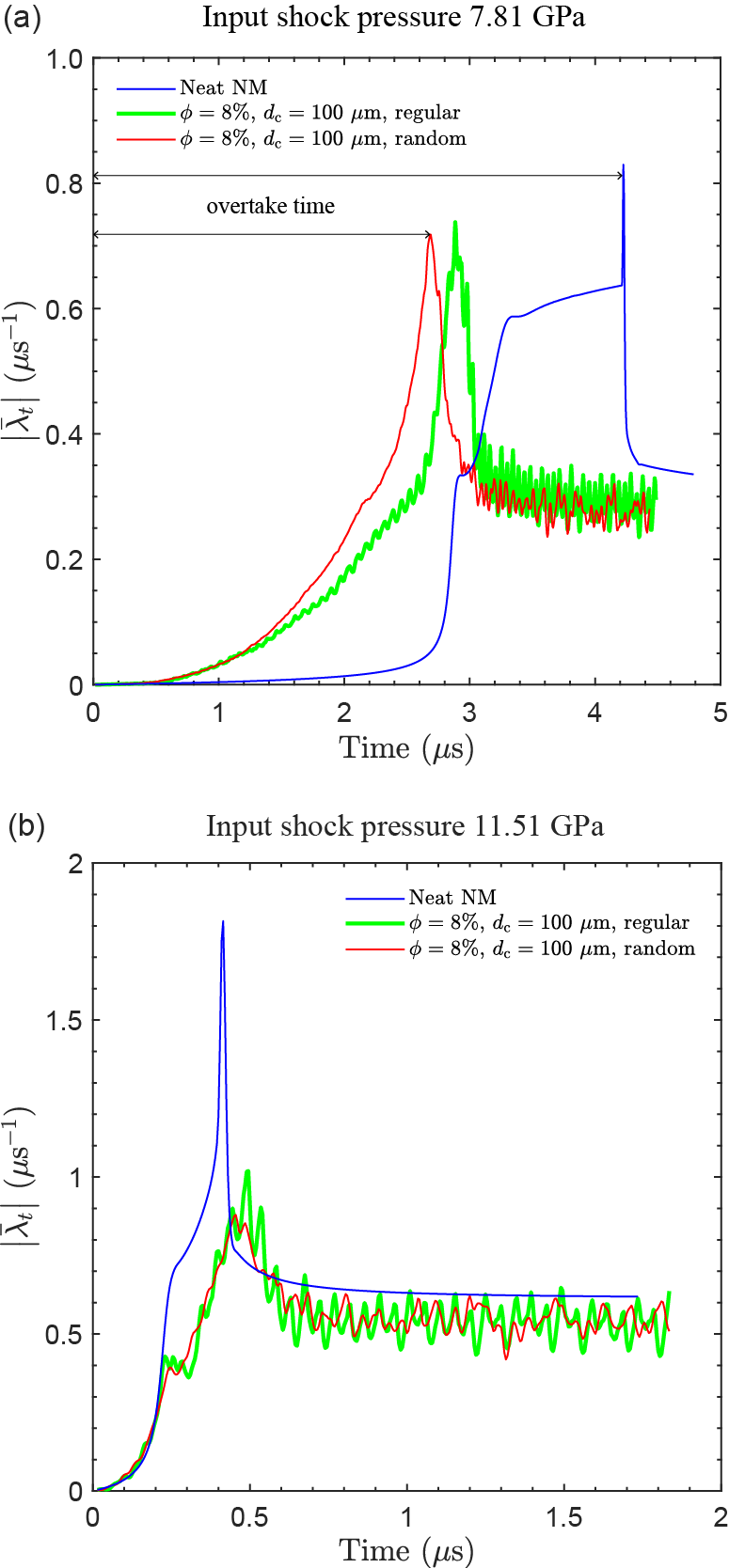}}
		\caption{The global reaction rate $\left| \bar{\lambda}_t \right|$ normalized with respect to the total mass of reactive NM, $M_{2,\mathrm{tot}}$, plotted as a function of time. The results compared here are for the scenarios of neat NM (blue curve), an array of regularly spaced cavities (green curve), and randomly distributed cavities (red curve) of $\phi=8\%$ and $d_\mathrm{c}=100~\mu\mathrm{m}$ subjected to an incident shock pressure of (a) $7.81 \mathrm{GPa}$ and (b) $11.51 \mathrm{GPa}$ .}
	\label{Fig6}
\end{figure}

The normalized global reaction rate as a function of time for the case of neat NM with an incident shock pressure of $7.81~\mathrm{GPa}$ is plotted as the blue curve in Fig.~\ref{Fig6}(a). As shown on this figure, for the case of neat NM, the reaction rate is of an insignificant value before $t=2.4~\mu\mathrm{s}$. A rapid increase in $\left| \bar{\lambda}_t \right|$ occurs at $t=2.6$-$2.8~\mu\mathrm{s}$. After $t=3.2~\mu\mathrm{s}$, this rapid increase slows down until approximately $t=4.2~\mu\mathrm{s}$ where the curve of $\left| \bar{\lambda}_t \right|$ abruptly increases to its peak value. After reaching its peak, $\left| \bar{\lambda}_t \right|$ undergoes an abrupt decrease. From $t=4.3~\mu\mathrm{s}$ onwards, the reaction rate keeps decreasing but at much slower pace until the end of the simulation at $t=4.8~\mu\mathrm{s}$. The time at which the curve of $\left| \bar{\lambda}_t \right|$ reaches its peak is measured as $t=4.22~\mu\mathrm{s}$ from Fig.~\ref{Fig6}, agreeing with the measurement of the time when the second shock wave overtakes the incident shock front from the $x$-$t$ diagram plotted in Fig.~\ref{Fig5}(a). Thus, the time that has elapsed since the incident shock enters the reactive NM (at $t=0$) until the global reaction rate reaching its peak value can be considered as the overtake time for the resulting SDT process.

The results of $\left| \bar{\lambda}_t \right|$ as a function of time for the cases with a regular array of cavities and randomly distributed cavities subjected to an incident shock pressure of $7.81~\mathrm{GPa}$ are plotted as the green and red curves in Fig.~\ref{Fig6}(a), respectively. For both of these cavity-laden scenarios, the resulting reaction rates start to significantly increase very shortly after the incident shock enters the reactive NM mixture. As shown in Fig.~\ref{Fig6}(a), the results of $\left| \bar{\lambda}_t \right|$ for the cases with regularly and randomly distributed cavities remain approximately the same until $t=1.2~\mu\mathrm{s}$, and separate afterward: The global reaction rate for the case with a random distribution increases more rapidly than that for a regular array of cavities. The curve of $\left| \bar{\lambda}_t \right|$ for a random distribution reaches its peak at $t=2.67~\mu\mathrm{s}$ while that for the case of a regular array peaks at $t=2.88~\mu\mathrm{s}$. These times are thus measured as the overtake times for the SDT processes occurring in cavity-laden NM mixtures. After attaining the peaks, both curves gradually decrease to a quasi-plateau from roughly $t=3.3~\mu\mathrm{s}$ onwards.

Fluctuations can be observed on the curve of $\left| \bar{\lambda}_t \right|$ for the case with a regular array of cavities (i.e., green curve in Fig.~\ref{Fig6}(a)) throughout the entire transition process captured by the simulation. These fluctuations are of an equal frequency, i.e., local peak-to-valley time interval. The amplitude of these fluctuations is greater after the global reaction rate decreases to the quasi-plateau value. The curve of $\left| \bar{\lambda}_t \right|$ for the case with randomly distributed cavities (i.e., red curve in Fig.~\ref{Fig6}(a)) does not exhibit any significant fluctuations until it decreases to its quasi-plateau from $t=3.3~\mu\mathrm{s}$ onwards.

For a significantly greater input shock pressure, $11.51~\mathrm{GPa}$, the results of $\left| \bar{\lambda}_t \right|$ for the case of neat NM and cavity-laden heterogeneous NM mixtures are plotted in Fig.~\ref{Fig6}(b). The curves corresponding to neat NM and heterogeneous mixtures are nearly the same up to $t=0.2~\mu\mathrm{s}$. The resulting $\left| \bar{\lambda}_t \right|$ from the neat NM case increases faster and reaches a greater peak value than those from the cavity-laden cases. After approximately $t=0.75~\mu\mathrm{s}$, the $\left| \bar{\lambda}_t \right|$ curves for all of the three cases have reached a plateau or a quasi-plateau with fluctuations. The plateau value for the case of neat NM is noticeably greater than the mean quasi-plateau value for the heterogeneous cases. The amplitudes of the fluctuations on the curve for the case of regularly spaced cavities are greater than those for the case with randomly distributed cavities.

In the experimental studies of SDT phenomena in heterogeneous explosives, the commonly used approach to determine overtake time is via fitting a function to the trajectory of the leading shock front (obtained from the gauge measurement of \textit{in-situ} particle velocity) and finding the time at which the shock acceleration is maximum, e.g., the approaches used by Hill and Gustavsen~\cite{Hill2002} and Gustavsen \textit{et al.}~\cite{Gustavsen2006}. A similar approach of finding the maximum shock acceleration is applied to selected cases of the current study. The measurement of overtake time via this approach closely agrees with that obtained via finding the time of maximum overall reaction rate described in this section. The comparison between these two approaches can be found in Appendix~\ref{append3}. The results of overtake time reported in the rest of this paper are obtained via the maximum reaction rate approach only.

\subsection{Overtake time vs. incident shock pressure}
\label{sec4_4}

The simulation results of the relationship between the overtake time and the input shock pressure for the scenarios with neat NM and regularly and randomly distributed cavities of $\phi=8\%$ and $d_\mathrm{c}=100~\mu\mathrm{m}$ are summarized in Fig.~\ref{Fig7} and plotted as blue circles, open green triangles, and solid red triangles, respectively. Note that the data are plotted with linear scales in this figure. For all of these three scenarios, the resulting overtake time decreases as incident shock pressure increases. For an incident shock pressure less than $9.42~\mathrm{GPa}$, the overtake time for both cavity-laden cases is less than that of the neat NM case; the overtake time for a random distribution of cavities is slightly, but still noticeably, less than that for the case with regularly spaced cavities. The difference among the resulting overtake times from these three scenarios becomes greater as input shock pressure decreases. For an input shock pressure greater than $10.05~\mathrm{GPa}$, the overtake times for all of the three scenarios are very close while those resulting from the cavity-laden NM mixtures are very slightly greater than that for neat NM.

\begin{figure}
\centerline{\includegraphics[width=0.43\textwidth]{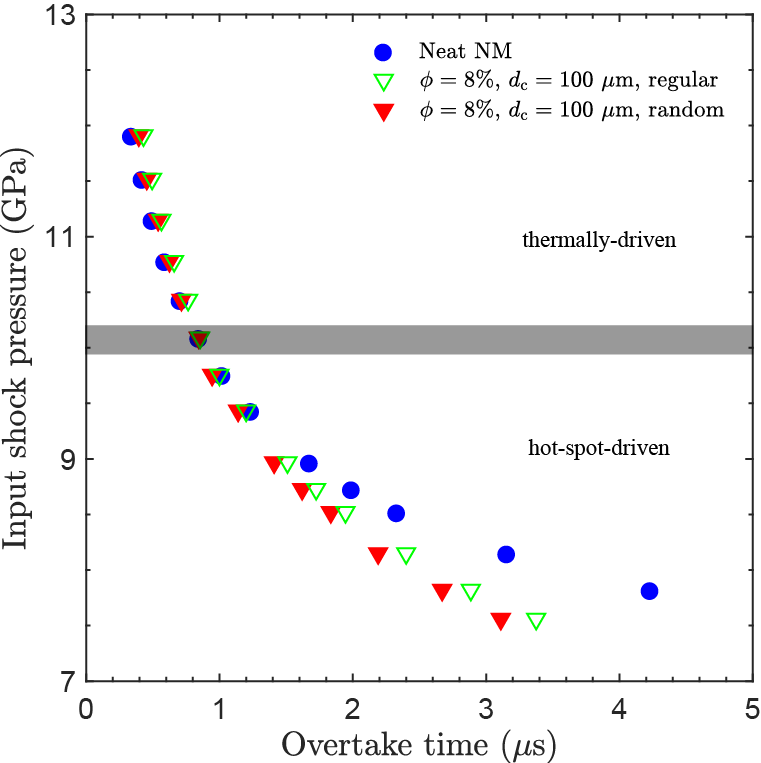}}
		\caption{Simulation results of overtake time as a function of incident shock pressure are plotted with linear scales. The results compared here are for the scenarios of neat NM (blue circles), an array of regularly spaced cavities (green open triangles), and randomly distributed cavities (red solid triangles)with $\phi=8\%$ and $d_\mathrm{c}=100~\mu\mathrm{m}$ .}
	\label{Fig7}
\end{figure}

The simulation results of overtake time vs. input shock pressure for the scenarios with neat NM and randomly distributed cavities of $\phi=8\%$ and $d_\mathrm{c}=100~\mu\mathrm{m}$ are replotted in Fig.~\ref{Fig8} to compare with experimental data of the SDT overtake time in both neat and GMB-sensitized NM~\cite{Campbell1961Liquid,Voskoboinikov1968,Lysne1973,Sheffield1989Report,Leal2000,Dattelbaum2009APS,Dattelbaum2010}. Note that Fig.~\ref{Fig8} is plotted with log-log scales--known as a ``Pop-plot''--which is how the experimental results of overtake times as a function of input shock pressure are commonly reported in the literature. The simulation results for neat NM are plotted as blue circles. The dashed line is a fitting line to the scatter of experimental data of overtake times reported in several independent studies.~\cite{Campbell1961Liquid,Voskoboinikov1968,Lysne1973,Sheffield1989Report,Leal2000} {The simulation results of the overtake times for neat NM qualitatively agree with the experimental data. The best quantitative agreement occurs over a range of input shock pressure from $8.14~\mathrm{GPa}$ to $9.42~\mathrm{GPa}$. The simulations result in greater overtake times than the experimental data for input shock pressures greater than $9.42~\mathrm{GPa}$. For input shock pressure less than $8.14~\mathrm{GPa}$, the simulation results of overtake time are slightly smaller than the experimental data.}

\begin{figure}
\centerline{\includegraphics[width=0.43\textwidth]{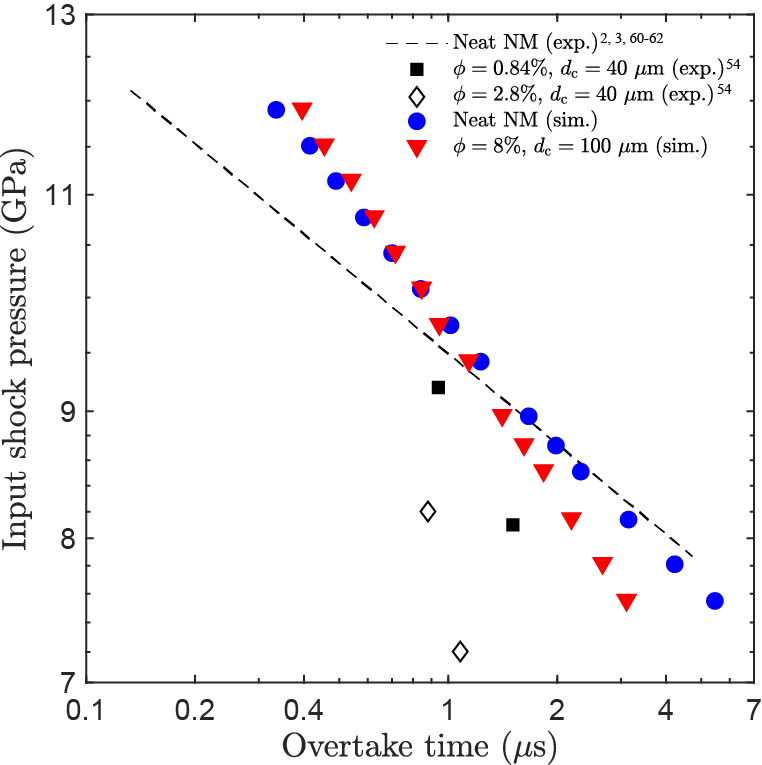}}
		\caption{Simulation results of overtake time as a function of incident shock pressure are presented as a Pop-plot (i.e., with log-log scales) to compare with the experimental results in the literature. The simulation results are for the scenarios of neat NM (blue circles) and randomly distributed cavities (red solid triangles) with $\phi=8\%$ and $d_\mathrm{c}=100~\mu\mathrm{m}$. The dashed line is a fitting line to the experimental data for neat NM.\cite{Campbell1961Liquid,Voskoboinikov1968,Lysne1973,Sheffield1989Report,Leal2000} The black squares and diamonds represent the experimental results for guar-gum-gelled NM mixed with GMBs of $d_\mathrm{c}=40~\mu\mathrm{m}$ and $\phi=0.84\%$ and $2.8\%$, respectively.\cite{Dattelbaum2010}}
	\label{Fig8}
\end{figure}

In Fig.~\ref{Fig8}, the simulation results of overtake time for the cases with randomly distributed cavities of $\phi=8\%$ and $d_\mathrm{c}=100~\mu\mathrm{m}$ are plotted as red triangles. The black squares and diamonds in Fig.~\ref{Fig8} represent the experimentally measured overtake times for SDT processes in guar-gum-gelled NM mixed with GMBs of $d_\mathrm{c}=40~\mu\mathrm{m}$ and $\phi=0.84\%$ and $2.8\%$, respectively. Note that only weight fractions of GMBs for these experimentally tested mixtures, i.e., $0.36~\mathrm{wt}\%$ and $1.2~\mathrm{wt}\%$, respectively, are provided in the original publication~\cite{Dattelbaum2010}; the corresponding volume fraction can be found in a later technical report~\cite{Dattelbaum2010Role} via examining the Scanning Electron Microscopy (SEM) images. In these experimental cases, the added GMBs are of lower porosities and a smaller size than those considered in the current simulations. The experimental results for GMB-sensitized NM mixtures demonstrate a greater reduction in overtake time from the results for neat NM than that resulting from the simulations. Possible reasons for the discrepancies between these simulation and experimental results (as shown in Fig.~\ref{Fig8}) will be discussed in Sect.~\ref{sec5_5}.

\subsection{Statistical analysis}
\label{sec4_5}

As revealed in a number of numerical studies, the hot spots formed over the course of a shock-induced collapse of a cavity are of irregular shapes; the distribution of temperature and density within these hot spots is spatially non-uniform. Thus, temperature and size of a hot spot are rather vaguely defined concepts since both of them vary over time and reflect neither how much material is present within such a hot spot nor how fast energy is released from it. To further quantitatively analyze the effect of hot spots based on meso-resolved simulation data, it is of importance to know the amount of material reacting at specific reaction rates throughout the SDT process. To this end, mass-weighted probability density functions (PDFs) over a broad range of volumetric rate of energy release (in the unit of $\mathrm{GW}/\mathrm{m}^3$ and denoted as $\dot{q}$) have been obtained via performing statistical analysis on the simulation data. The procedure to obtain these PDFs is described in detail in the following paragraphs.

At each time step, the two-dimensional distribution of all flow and state variables can be obtained directly from the simulations. {For each computational cell, the corresponding volumetric rate of energy release can be calculated as follows,}
\begin{equation}
\dot{q} = -z_2 \rho_2 K Q
\end{equation}
which is the right-hand side of the energy balance equation in Eq.~\ref{Eq1}. Via scanning through the simulation data of $\dot{q}$ over the entire domain at specific time $t$, a statistical counting of how many computational cells that are associated with a $\dot{q}$ in a specific interval can then be performed. Since the values of $\dot{q}$ of interest spans over a broad range, i.e., over nearly six orders of magnitude, this range of interest is equally divided on a decadic logarithmic scale. Thus, each specific interval of $\dot{q}$ over which the statistical counting is performed can be expressed as 
\begin{equation}
10^n~\mathrm{GW}/\mathrm{m}^3 \leq \dot{q} < 10^{n+\Delta n}~\mathrm{GW}/\mathrm{m}^3
\end{equation} 
where $\Delta n$ is the value of the equal interval on a decadic logarithmic scale of $\dot{q}$. To obtain all of the PDFs reported in this paper, $\Delta n = 0.02$ was used. The area under each of these PDF curves can be interpreted as the proportion of the initial total mass of reactive NM releasing energy at volumetric rates within a specific interval of $\dot{q}$ at time $t$.

To take the mass of material reacting at a specific rate into account, the statistical counting is performed on a mass-weighted basis. As one scans through the computational cells one by one, $m(\dot{q},t)$, the probability mass measuring the cumulative occurrence of cells with its local $\dot{q}$ falling within a specific interval at a specific time $t$, can be counted by adding the mass of NM occupying the volume of this cell, i.e., $z_2(x,y,t) \rho_2(x,y,t)$. After counting the probability mass $m(\dot{q},t)$ over a range of $\dot{q}$ from $10^2$ to $10^{10}~\mathrm{GW}/\mathrm{m}^3$ throughout the entire domain, the mass-weighted probability density function of $\dot{q}$ can be calculated as follows,
\begin{equation}
f(\dot{q},t) = \frac{m(\dot{q},t)}{M_{2,\mathrm{tot}} \Delta n}
\end{equation}
Note that $f(\dot{q},t)$ is based on the normalization with respect to the initial total mass of reactive NM so that the results of simulations with the same domain length but different domain widths can be compared on a common basis; the area under a PDF curve at a specific time is less than one. Since the material in the initially shocked zone and coming into the domain through the left boundary is set to be inert (i.e., product of NM with $\lambda=0$), it does not contribute to the statistical counting of the mass of reacting NM.

\begin{figure*}
\centerline{\includegraphics[width=0.8\textwidth]{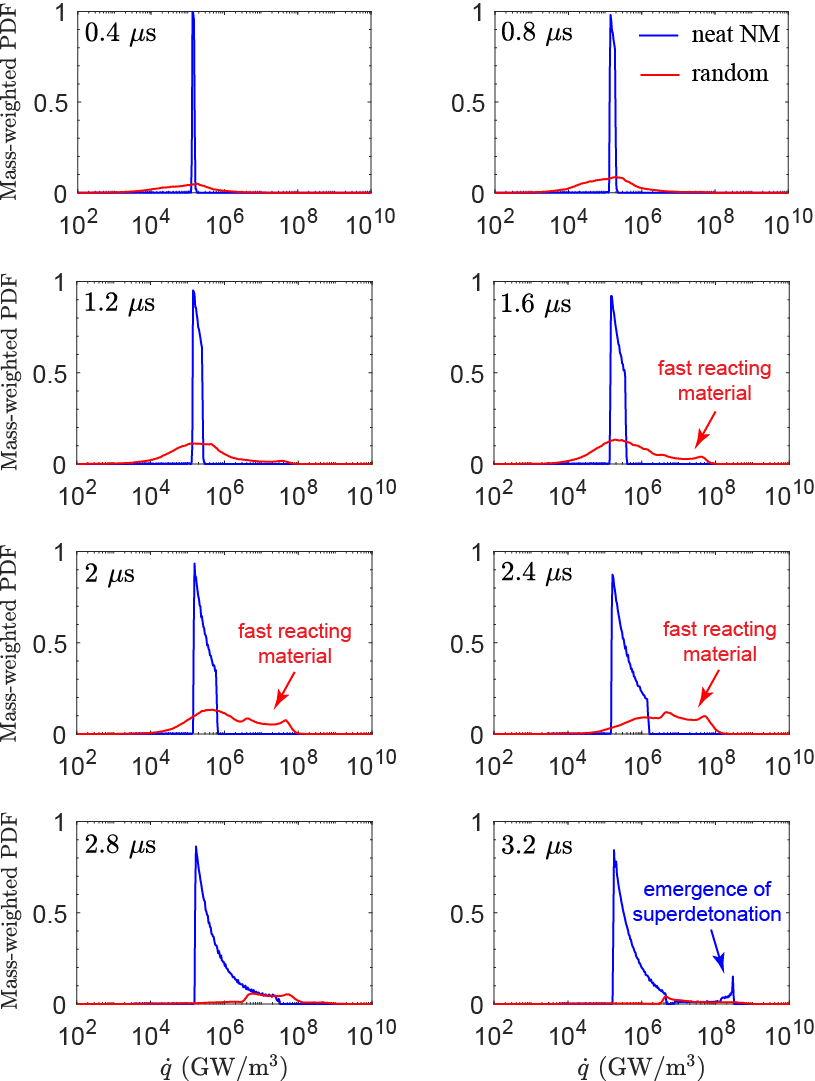}}
		\caption{Mass-weighted probability density functions $f(\dot{q},t)$ of volumetric rate of energy release $\dot{q}$ at various times for the cases of neat NM (blue curves, case shown in Fig.~\ref{Fig2}) and a mixture of NM and randomly distributed cavities of $\phi=8\%$ and $d_\mathrm{c}=100~\mu\mathrm{m}$ (red curves, case shown in Fig.~\ref{Fig4}) subjected to an incident shock pressure of $7.81~\mathrm{GPa}$.}
	\label{Fig9}
\end{figure*}

The mass-weighted PDFs of volumetric rate of energy release $\dot{q}$ for the cases of neat NM and a heterogeneous NM mixture with randomly distributed of cavities of $\phi=8\%$ and $d_\mathrm{c}=100~\mu\mathrm{m}$ are compared in Fig.~\ref{Fig9}. Each subfigure of Fig.~\ref{Fig9} is at a specific time which is marked at the top-left corner of the plot. The results shown in this figure are over the time from $t=0.4~\mu\mathrm{s}$ to $3.2~\mu\mathrm{s}$. The PDFs for neat NM and the case with randomly distributed cavities are plotted as blue and red curves, respectively. At $t=0.4~\mu\mathrm{s}$, the PDF curve for neat NM exhibits a sharp peak of nearly $f=1$ located at a narrow range slightly above $10^5~\mathrm{GW}/\mathrm{m}^3$ and $f=0$ elsewhere. The PDF curve for the heterogeneous case at $t=0.4~\mu\mathrm{s}$ is spread out over a range from approximately $\dot{q}=10^{4}~\mathrm{GW}/\mathrm{m}^3$ to $10^{6}~\mathrm{GW}/\mathrm{m}^3$. Without a sharp peak, the PDF curve for the heterogeneous case exhibits a smooth ``hump'' with its maximum probability density at approximately the same $\dot{q}$ where the high peak of the PDF curve for neat NM is located. As time progresses from $t=0.4~\mu\mathrm{s}$ to $2.8~\mu\mathrm{s}$, the PDF curve for the neat scenario gradually expands into the region of greater $\dot{q}$ up to above $10^{7}~\mathrm{GW}/\mathrm{m}^3$; the peak value of $f$ decreases and remains at roughly $\dot{q}=10^{5}~\mathrm{GW}/\mathrm{m}^3$.  For the heterogeneous case, from $t=0.4~\mu\mathrm{s}$ to $1.8~\mu\mathrm{s}$, the PDF reaches its maximum at roughly the same $\dot{q}$ as where the peak for the neat NM curve is; the area under the PDF curve, as indicated by the red arrows, significantly increases in the range of greater $\dot{q}$ from roughly $10^{6}~\mathrm{GW}/\mathrm{m}^3$ to $10^{8}~\mathrm{GW}/\mathrm{m}^3$ over which the PDF for neat NM remains zero. The PDF curve for the heterogeneous case has its hump shifted away from $\dot{q}=10^{5}~\mathrm{GW}/\mathrm{m}^3$ after $t=2.4~\mu\mathrm{s}$ onwards, and eventually at $t=3.2~\mu\mathrm{s}$, it evolves to a much smaller hump located at roughly $\dot{q}=10^{6}~\mathrm{GW}/\mathrm{m}^3$. For the case with neat NM at $t=3.2~\mu\mathrm{s}$, a second peak appears in the region where $\dot{q} \geq 10^{8}~\mathrm{GW}/\mathrm{m}^3$ as indicated by the blue arrow.

\begin{figure*}
\centerline{\includegraphics[width=0.8\textwidth]{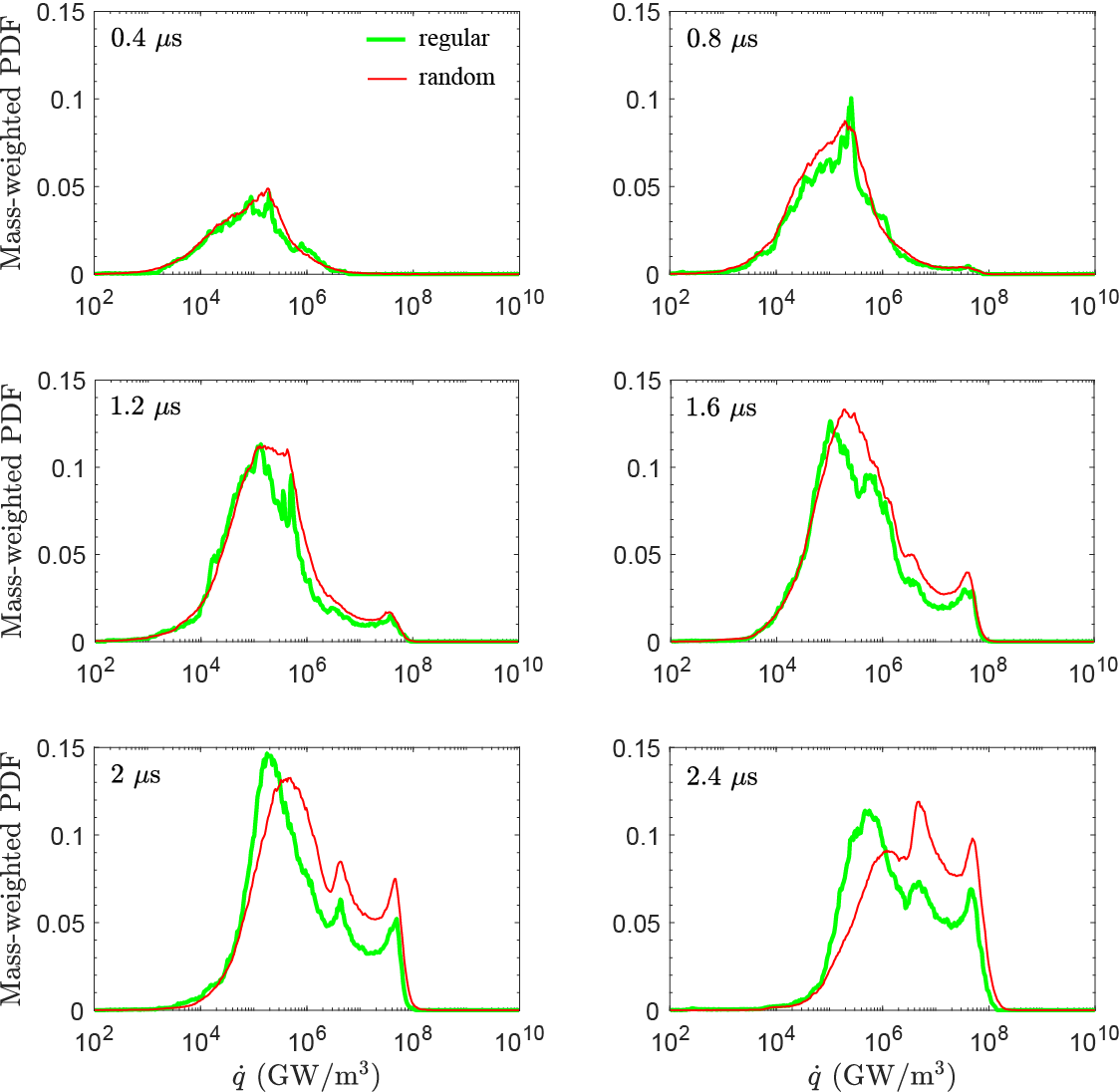}}
		\caption{Mass-weighted probability density functions $f(\dot{q},t)$ of volumetric rate of energy release $\dot{q}$ at various times for the cases of NM mixed with an array of regularly spaced cavities (green curves, case shown in Fig.~\ref{Fig3}) and randomly distributed cavities (red curves, case shown in Fig.~\ref{Fig4}) of $\phi=8\%$ and $d_\mathrm{c}=100~\mu\mathrm{m}$ subjected to an incident shock pressure of $7.81~\mathrm{GPa}$.}
	\label{Fig10}
\end{figure*}

In Fig.~\ref{Fig10}, the mass-weighted PDFs of $\dot{q}$ for the scenarios with an array of regularly spaced cavities (plotted as green curves) and randomly distributed cavities (plotted as red curves) of $\phi=8\%$ and $d_\mathrm{c}=100~\mu\mathrm{m}$ are compared. Note that the results for the case with a random distribution is the same as that shown in Fig.~\ref{Fig9} but plotted with a smaller scale of $f(\dot{q},t)$, i.e., from $0$ to $0.15$ in this figure instead of from $0$ to $1$ in Fig.~\ref{Fig9}, in order to more clearly demonstrate the difference between the cases with a regular and random distributions. As shown in the two subfigures for $t=0.4~\mu\mathrm{s}$ and $0.8~\mu\mathrm{s}$, the PDF curves for regular and random cases have roughly the same overall behavior while the curve for the random case appears to be relatively smoother and the one for the regular case exhibits more small-amplitude fluctuations. From $t=1.2~\mu\mathrm{s}$ onwards, the area under both curves increases over the range of $\dot{q}$ from roughly $10^{6}~\mathrm{GW}/\mathrm{m}^3$ to $10^{8}~\mathrm{GW}/\mathrm{m}^3$. This increase in area over the range of greater $\dot{q}$ is faster for the case with randomly distributed cavities than that for the case with a regular distribution.

\section{Discussion}
\label{sec5}

The meso-resolved simulation results and post-simulation analysis presented in the previous section reveal various facets of the SDT processes in neat NM and cavity-laden mixtures of NM. In this section, these results are interpreted to elucidate the key mechanisms underlying the shock-initiation behaviors of different scenarios. The findings from the current simulation results are compared with those speculated or experimentally discovered in previous studies. Possible reasons for the discrepancy between the simulation results and experimental data of the SDT overtake times are discussed. 

\subsection{Neat NM}
\label{sec5_1}

The simulations for the case of neat NM performed in this study captured an SDT mechanism featuring a sudden thermal explosion after an induction period since the passage of the incident shock and a gradually developed superdetonation eventually overtaking the incident shock wave. {All of these key features (as indicated on the $x$-$t$ diagram shown in Fig.~\ref{Fig5}(a) and the time histories of particle velocity shown in Fig.~\ref{Fig_u_neat}) agree with the SDT mechanism in neat NM elucidated in the literature.\cite{Holland1957,Cook1958,Chaiken1960,Campbell1961Liquid,Sheffield1989Report}.} {As shown in Fig.~\ref{Fig5}(a), the thermal explosion occurs at the inflow material interface. The left-propagating blast wave emanating from the explosion propagates through the inflow material and increases the density in this region, which can be observed in the snapshots from $t=2.7~\mathrm{\mu s}$ onwards.} It can be noticed from Figs.~\ref{Fig5}(a), \ref{Fig_u_neat}, and Fig.~\ref{Fig6}(a) that, following the thermal explosion, the reaction wave develops into a superdetonation over a finite amount of time from approximately $t=2.8~\mu\mathrm{s}$ to $3.3~\mu\mathrm{s}$. This gradual development of a superdetonation in neat NM better agrees with that experimentally observed by Sheffield \textit{et al}. via \textit{in-situ} measurement of particle velocity\cite{Sheffield1989Report}, rather than Chaiken and Campbell \textit{et al}.'s speculation~\cite{Chaiken1960,Campbell1961Liquid} that a steady superdetonation is immediately initiated by the thermal explosion. Further, it is of interest to note that the history of the global reaction rate for neat NM shown in Fig.~\ref{Fig6}(a) exhibits a qualitatively similar behavior as that of the temporal profile of the radiance emitted by neat NM undergoing an SDT process, which was recorded by Bouyer~\textit{et al}. for spectral analysis to determine the corresponding temperature profile (as shown in Fig.~4 of this paper)~\cite{Bouyer2006CF}.

\subsection{Cavity-laden heterogeneous mixtures of NM}
\label{sec5_2}

As shown by Fig.~\ref{Fig5}(b) and (c), the $x$-$t$ diagram of $y$-direction averaged pressure contour plot for the case of NM mixed with air-filled cavities, the mechanism governing the SDT is completely different from that for the case of neat NM. The pressure behind the leading shock gradually increases, resulting in a smooth transition from an incident shock to a quasi-steady detonation wave. {No evidence of thermal explosion or superdetonation can be found in the $x$-$t$ diagram or the time histories of particle velocities (Fig.~\ref{Fig_u_random}) for heterogeneous NM subjected to an incident shock pressure of $7.81~\mathrm{GPa}$. This difference can also be found by comparing the time histories of the global reaction rate for neat NM and cavity-laden heterogeneous mixtures of NM in Fig.~\ref{Fig6}(a). The temporal profiles of $\left| \bar{\lambda}_t \right|$ for the heterogeneous NM mixtures (i.e., red and green curves) start to gradually increase from a very early stage, at approximately $t=0.4~\mu\mathrm{s}$, without undergoing a long induction process. These analyses show that significant energy release is triggered immediately downstream from the leading shock and is gradually accelerated over time, which is consistent with the experimental findings for SDT processes in NM mixed with silica beads and GMBs~\cite{Dattelbaum2009APS,Dattelbaum2010} and heterogeneous solid explosives~\cite{Ramsay1965}.}

This enhancement in SDT sensitivity has long been attributed to the formation and growth of hot spots. As revealed in a number of computational studies, the hot spots formed due to shock--heterogeneity interactions are of irregular shapes and sizes.~\cite{Ball2000,Swantek2010AIAA,Swantek2010JFM,Hawker2012,Lauer2012,Ozlem2012,Michael2014DS,Kapila2015,Betney2015,Apazidis2016,Michael2018_I,Michael2018_II,Michall2019Book} In addition, as shown in this study (e.g., Fig.~\ref{Fig4}) and others, the wave interaction due to the collapse of neighboring cavities may further complicate the sensitizing mechanisms of hot spots. Concerning these complexities, using hot-spot number density derived from the volume or weight fraction of the heterogeneous inclusions to quantitatively gauge the effect of hot spot is likely an oversimplification. Taking the advantage of the full flow-field information obtained from the meso-scale simulations, statistical analysis has been performed to reveal the mass spectrum (i.e., mass-weighted PDF) over a broad range in reaction rate at different stages in an SDT process.

For an input shock pressure of $7.81~\mathrm{GPa}$, as shown in Fig.~\ref{Fig9}, the area of the PDF curve  for neat NM (blue curve) is initially concentrated at a relatively low reaction rate; the gradual expansion of  this curve towards the direction of greater reaction rates indicates the increase in the mass of shocked material. In contrast, the PDF curve for the heterogeneous case (red curve) exhibits a hump-shape and is spread out over a wider range in reaction rate since a very early stage. The amount of mass at greater reaction rates beyond the span of the neat NM curve is likely associated with hot spots. The tail of the heterogeneous PDF curves into the region of lower reaction rates is attributed to the remaining amount of reacting material with smaller mass fractions of reactant $\lambda$. A continuously increasing amount of fast reacting material, as indicated by the red arrows in Fig.~\ref{Fig9}, further accelerates the initiation process, demonstrating the collective effect of a large number of hot spots regardless of their shapes and sizes.

\subsection{Effect of input shock pressure}
\label{sec5_3}
As shown in Fig.~\ref{Fig7}, the overtake times for the heterogeneous cases are significantly shorter than those for the cases with neat NM subjected to incident shock pressures less than $9.42~\mathrm{GPa}$; for an incident shock pressure greater than this value, the overtake time for neat NM is slightly shorter than those for heterogeneous NM mixtures. For a cavity-laden NM mixture subjected to a relatively weak input shock, as revealed by the simulation results and statistical analysis shown in Figs.~\ref{Fig6}(a) and \ref{Fig9}, the initiation process is dominated by the mechanism of hot spots. The resulting SDT behaviors are thus considered to be in the hot-spot-driven regimes.~\cite{Dattelbaum2010Role} For increasingly weaker input shock strength, the sensitizing effect of hot spots is more pronounced as shown in Figs.~\ref{Fig7} and \ref{Fig8}. {For further lower input shock pressures, shock focusing and reflection may not be able to induce sufficiently high temperature for fast reaction, and viscous heating is speculated to dominate the formation and growth of hot spots.\cite{Field1992hot} It is thus of importance to incorporate material viscosity into the model to examine the SDT behavior under low input shock pressures.}

If the strength of the input shock is greater than a certain value, given an elevated post-shock temperature, the induction process is very short for both neat NM and heterogeneous mixtures. As shown in Fig.~\ref{Fig6}(b), the global reaction rate for both neat NM and heterogeneous mixtures start to significantly increase immediately after the incident shock entering the reactive material and remain nearly the same at an early stage of the SDT process. These results suggest that the initiation process following a strong input shock is dominated by the energy release in the bulk material, rather than at localized high-temperature regions. Such SDT behaviors are considered to be in the thermally-driven regimes wherein the energetically diluting effect of adding cavities dominates over the sensitizing effect via producing hot spots.~\cite{Dattelbaum2010Role} Thus, for greater input shock pressures, the subsequent global reaction rates for the heterogeneous cases are lower than that for neat NM (as shown in Fig.~\ref{Fig6}(b)); the overtake times resulting from heterogeneous cases are slightly longer than those for neat NM (as shown in Fig.~\ref{Fig7}). For the selected values of porosity and cavity size (i.e., $\phi=8\%$ and $d_\mathrm{c}=100~\mu\mathrm{m}$), an input shock pressure of approximately $9.42~\mathrm{GPa}$, as indicated by the gray band in Fig.~\ref{Fig7}, marks the balance between the sensitizing and energetically diluting effects of the addition of cavities, or the boundary between hot-spot-driven and thermally-driven SDT regimes. The cross-over from a hot-spot-driven regime to a thermally-driven regime for an increasingly greater input shock strength has been experimentally identified from the SDT tests with gelled NM mixed with silica beads of $6\%$ mass fraction and $40~\mu\mathrm{m}$ diameter.~\cite{Dattelbaum2009APS,Dattelbaum2010}

\subsection{Regular vs. random distributions of cavities}
\label{sec5_4}

For the same porosity and size of the cavities, in hot-spot-driven regimes, the overtake times resulting from the cases with randomly distributed cavities are shorter than those from the cases with a regular distribution as shown in Fig.~\ref{Fig7}(a). Comparing the mass spectrum over reaction rate between the cases with random and regular distributions for an incident shock pressure of $7.81~\mathrm{GPa}$ as shown in Fig.~\ref{Fig10}, it can be noticed that, at $t=1.2~\mu\mathrm{s}$, the random distribution starts to result in a greater amount of material reacting at rates above $\dot{q}=10^{6}~\mathrm{GW}/\mathrm{m}^3$ than that for the regular case. Due to a thermally positive feedback mechanism, the amount of fast reacting material increases faster in the cases with a random distribution, and thus, sooner leads to the detonation overtake.

The further enhancement in shock-initiation sensitivity via randomizing the spatial distribution of cavities might be due to the fact that there are local regions with more closely spaced cavities, i.e., the spacing between neighboring cavities is smaller than the average spacing $\delta_\mathrm{c}$ calculated via Eq.~\ref{Eq6_1}. As the incident shock passes through such regions, the local number density of hot spots is greater than the average value. Thus, a local cluster of hot spots is formed. The localized explosions triggered by this cluster of hot spots may interact with each other via emanating strong pressure waves. This collective behavior of clustering hot spots may further enhance the energy release rate within the cluster and accelerate the overall SDT process via a thermally positive feedback mechanism. 

\subsection{Simulation results vs. experimental data}
\label{sec5_5}

As shown in Fig.~\ref{Fig8}, the simulation results of overtake times for neat NM agree fairly well with the experimental data for relatively low input shock pressures over a range from $7.55~\mathrm{GPa}$ to $9.42~\mathrm{GPa}$, but exhibit an increasingly larger deviation as input shock pressure increases beyond $9.42~\mathrm{GPa}$. The agreement and discrepancy between simulation results and experimental data for neat NM reflect the fact that, in the reaction rate model adapted for these simulations (Eq.~\ref{Eq6}), the value of the pre-exponential factor $C$ is calibrated based on the experimental measurement of overtake time of liquid NM for shock pressures ranging from $7.5$ to $9.5~\mathrm{GPa}$, and the value for $T_\mathrm{a}$ was calibrated based on the \textit{in-situ} gauge data of particle velocity for an input shock pressure of $9.1~\mathrm{GPa}$. 

{
The choice and simplifications made to the EoS of NM and its products also impose major limitations to the quantitative accuracy of the model. As Menikoff and Shaw pointed out, the choice of EoS and reaction rate model is crucial for an accurate modeling of detonation waves in NM.\cite{Menikoff2011} Using the same Cochran-Chan EoS (Eqs.~\ref{Eq2}-\ref{Eq4}) to govern the reactant and products of liquid NM does not adequately describe the physics in the reaction zone and the subsequent flow expansion. In future efforts, following the approach proposed by Wilkinson \textit{et al}.~\cite{Wilkinson2017}, the EoS for NM reactant and reaction rate model can be calibrated using the up-to-date gauge measurement data, and the product EoS can be obtained by fitting to the principal adiabat calculated by an ideal detonation code such as \textit{IDex}. Simulations treating the reactant and product of an explosive as two miscible materials governed by different EoS can be implemented using the \textit{MiNi16} formulation~\cite{Michael2016JCP,Michael2018JCP}. In addition, considering the constant-volume heat capacity and Gr\"{u}neisen coefficient for NM to be constant values may also introduce inaccuracies to the simulation results. The thermodynamically consistent EoS developed by Winey~\textit{et al}.\cite{Winey2000} for shocked liquid NM with density- and temperature-dependent heat capacity, thermal pressure coefficient, isothermal bulk modulus, and Gr\"{u}neisen coefficient can be incorporated in further development of the model.
}

For the cases under hot-spot-driven regimes, the (perhaps only) available experimentally measured overtake times for SDT in gelled NM mixed with GMBs of $d_\mathrm{c}=40~\mu\mathrm{m}$ and $\phi=0.84\%$ and $2.8\%$ are significantly less than those resulted from the current simulations with cavities of $\phi=8\%$ and $d_\mathrm{c}=100~\mu\mathrm{m}$. This difference can be attributed to several factors: (1) The overall volumetric number density of cavities; (2) clustering effect of hot spots due to a spatially non-uniform distribution of cavities; (3) the discrepancy in hot-spot temperature arising from two and three dimensions. Each factor is briefly discussed in the following paragraphs.

In a three-dimensional system, knowing the size and porosity of the cavities in the mixture, the volumetric number density can be calculated as 
\begin{equation}
\rho_\mathrm{N} = \frac{6\phi}{\pi {d_\mathrm{c}}^3}
\end{equation}
For the two experimentally tested mixtures with GMBs of $d_\mathrm{c}=40~\mu\mathrm{m}$ and $\phi=0.84\%$ and $2.8\%$, the corresponding number densities are $\rho_\mathrm{N} \approx 250~{\mathrm{mm}}^{-3}$ and $835~{\mathrm{mm}}^{-3}$, respectively. For the heterogeneous case considered in the simulation results in Fig.~\ref{Fig8}, the equivalent three-dimensional number density of cavities is determined to be $\rho_\mathrm{N} \approx 153~{\mathrm{mm}}^{-3}$. Since the $\rho_\mathrm{N}$ of cavities for the simulation cases is smaller than the $\rho_\mathrm{N}$ of GMBs in the experimentally tested mixtures, the hot-spot sensitizing effect on the SDT process is likely less significant in the current simulation cases.

As discussed in Sect.~\ref{sec5_5}, the overall number density of cavities or hot spots may not be the sole parameter that determines the resulting SDT behaviors. The clustering of cavities can likely enhance their sensitizing effect. It is of importance to recall that, in the current simulations, a minimum cavity spacing of $2 d_\mathrm{c}$ is imposed to the random distribution of cavities. For the cases with cavities of $\phi=8\%$ and $d_\mathrm{c}=100~\mu\mathrm{m}$, this imposed minimum cavity spacing is $250~\mu\mathrm{m}$ while the average spacing is $313~\mu\mathrm{m}$, which renders the random distribution in these simulations rather uniform over space. Likely due to this spatial uniformity of the random distribution of cavities, the clustering effect of hot spots is less pronounced in the simulations than that in the experimentally tested mixtures. In future studies, it might be of interest to impose some controllable spatial non-uniformities to the random distribution of meso-scale heterogeneities and probe the effect of such non-uniformities on the SDT behavior of the explosive mixtures.

The shock-induced collapse of a single air-filled cavity in liquid NM has been investigated by Michael and Nikiforakis via both two- and three-dimensional simulations using the same set of governing equations as that in the current work.\cite{Michael2018_I,Michael2018_II} These authors have demonstrated that the peak temperature reached by a hot spot in a three-dimensional configuration can be nearly $1.5$ times of the peak hot-spot temperature resulting from a two-dimensional scenario (as shown in Fig.~$11$ of Ref.~[28]). This difference in hot-spot temperature can lead to an order-of-magnitude greater reaction rate at the kernel of a hot spot formed during the shock-induced collapse of a spherical cavity in three dimensions. Hence, in the current two-dimensional simulations, the hot-spot effect is likely less significant than in realistic three-dimensional scenarios.

{Further, numerical inaccuracies may be rooted in the diffuse-interface approach to separate the two fluids. A cavity collapses due to the passage of a shock wave, and the initially sharp interface grows into an artificial mixing layer during this process. As pointed out by Ozlem \textit{et al}.\cite{Ozlem2012} and Michael and Nikiforakis\cite{Michael2018_II}, the thermodynamics in this artificial mixing layer are open to questions. Since the thickness of artificial mixing layer decreases for an increasingly finer grid resolution, the numerical convergence of the SDT overtake time results demonstrated in Appendix A indicates that the results are not significantly affected by the numerical uncertainty arising from the artificial mixing layers. To further examine this issue, it is of interest to compare the current results to those obtained from simulations using a sharp-interface approach, such as the level-set methods used to study void collapse in solid explosives\cite{Sambasivan2013,Kapahi2013,Rai2015JAP,Rai2017PRF,Rai2018JAP}.}

\section{Concluding remarks}
\label{sec6}

The complete SDT process in a heterogeneous NM mixture that contains a statistically significant number of air-filled cavities has been studied via GPU-enabled computational simulations. Without invoking any empirically calibrated, phenomenological reaction models, the characteristic shock-induced initiation process in such a heterogeneous explosive mixture can be captured by these meso-resolved simulations. {The SDT mechanisms revealed by the simulations for neat NM and cavity-laden NM mixtures are qualitatively consistent with those hypothesized based upon the experimental evidence.} The simulation results also show that the sensitizing effect of the addition of heterogeneities is significant for a relatively weak input shock wave; for a sufficiently strong input shock wave, the presence of cavities slightly impedes the initiation process due to a dilution of the energetic medium. A uniformly random distribution of cavities has a noticeably more pronounced sensitizing effect on the shock-induced initiation of the mixture than a regular array of cavities does. With the detailed flow-field data obtained from the meso-resolved simulations, a framework of statistical analysis has been proposed to quantitatively examine the hot-spot mechanisms underlying the SDT process. Further development towards a quantitatively better agreement between the simulation and experimental results will be carried out in future efforts. 

\begin{acknowledgments}
XCM is supported by an NSERC Postdoctoral Fellowship (PDF-502505-2017). Computing resources used in this work were provided by Compute Canada and University of Cambridge. The authors are grateful to J. Loiseau for useful discussions in developing this paper.
\end{acknowledgments}

\appendix

\section{Numerical convergence study}
\label{append1}

In this paper, the key result obtained from each SDT simulation is the detonation overtake time. Thus, numerical convergence tests have been performed on this time scale, which characterizes the resulting SDT behavior. Since the cavity diameter is chosen to be $d_\mathrm{c} = 100~\mu\mathrm{m}$ for all cavity-laden cases reported in this paper, this diameter is used for the convergence tests.

For the case with an array of regularly spaced cavities of $\phi=8\%$ and $d_\mathrm{c}=100~\mu\mathrm{m}$ subjected to an input shock pressure of $7.81~\mathrm{GPa}$, the normalized global reaction rate $\left| \bar{\lambda}_t \right|$ (calculated via Eqs.~\ref{Eq10} and \ref{Eq11}) as a function of time resulting from simulations with various computational grid sizes are compared in Fig.~\ref{FigA1}. The overtake times obtained from this set of convergence tests for a regular distribution of cavities are summarized in Table~\ref{Table_A1}. The temporal profile of $\left| \bar{\lambda}_t \right|$ and the result of overtake time do not change significantly as the computational grid size is further reduced beyond $\mathrm{d}x=1~\mu\mathrm{m}$, i.e., $100$ grids across the diameter of a cavity. For this selected case, the length and width of the computational domain are $24~\mathrm{mm}$ and $0.313~\mathrm{mm}$, respectively. At the highest numerical resolution reported in this study, i.e., $\mathrm{d}x=0.5~\mu\mathrm{m}$, there are $48000 \times 626$ computational points in the domain. To obtain most of the results of overtake times, the simulations were performed at the second highest resolution, i.e., $\mathrm{d}x=1\mu\mathrm{m}$, with $24000 \times 313$ computational points in the domain.

\begin{figure}
\centerline{\includegraphics[width=0.43\textwidth]{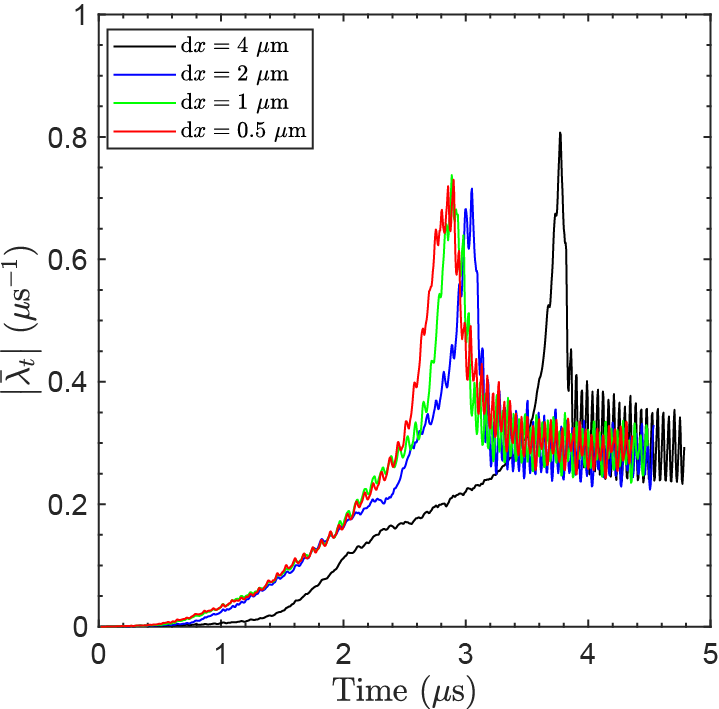}}
		\caption{Numerical convergence test: The normalized global reaction rate $\left| \bar{\lambda}_t \right|$ as a function of time for various computational grid sizes for the case with an array of regularly spaced cavities of $\phi=8\%$ and $d_\mathrm{c}=100~\mu\mathrm{m}$ subjected to an input shock pressure of $7.81~\mathrm{GPa}$.}
	\label{FigA1}
\end{figure}

\begin{table}[ht]
\begin{center}
\caption{Numerical convergence tests for the case with an array of regularly spaced cavities of $\phi=8\%$ and $d_\mathrm{c}=100~\mu\mathrm{m}$ subjected to an input shock pressure of $7.81~\mathrm{GPa}$.}
\label{Table_A1}
\begin{tabular}{| c | c | c | c | c |}
\hline
$\mathrm{d}x$ & $4~\mu\mathrm{m}$ & $2~\mu\mathrm{m}$ & $1~\mu\mathrm{m}$ & $0.5~\mu\mathrm{m}$ \\
\hline
Overtake time & $3.77~\mu\mathrm{s}$ & $3.05~\mu\mathrm{s}$ & $2.88~\mu\mathrm{s}$ & $2.88~\mu\mathrm{s}$\\
\hline
\end{tabular}
\end{center} 
\end{table}

For the case with randomly distributed cavities of $\phi=5\%$ and $d_\mathrm{c}=100~\mu\mathrm{m}$ subjected to an input shock pressure of $7.81~\mathrm{GPa}$, the normalized global reaction rate $\left| \bar{\lambda}_t \right|$ as a function of time resulting from simulations with various computational grid sizes are compared in Fig.~\ref{FigA2}. The overtake times obtained from this set of convergence tests for a regular distribution of cavities are summarized in Table~\ref{Table_A2}. Refining the computational grid from $\mathrm{d}x=2~\mu\mathrm{m}$ to $1~\mu\mathrm{m}$ results in a much smaller change in overtake time than that resulting from a grid refinement from $\mathrm{d}x=4~\mu\mathrm{m}$ to $2~\mu\mathrm{m}$. As previously shown for the case with a regular distribution, further reducing $\mathrm{d}x$ below $1~\mu\mathrm{m}$ would not lead to a significant change in the global overtake time. Thus, all of the results for random distributions of cavities reported in this paper were obtained from simulations performed at a numerical resolution of $\mathrm{d}x=1~\mu\mathrm{m}$. Note that, as the transverse width of the computational domain $W$ has to be sufficiently large so that the overall resulting dynamics for randomly distributed cavities are statistically converged (as discussed in Appendix~\ref{append2}). For the current case with an input shock pressure of $7.81~\mathrm{GPa}$, the length and width of the computational domain are $24~\mathrm{mm}$ and $2~\mathrm{mm}$, respectively. There are thus $24000 \times 2000$ computational points in the domain. Given the current model formulation, the memory requirement of such a simulation is approximately $5~\mathrm{GB}$. Given a $16$-$\mathrm{GB}$ NVIDIA Tesla P100 GPU, it is unlikely feasible to double the numerical resolution for the same domain size. The use of multiple-GPU computing via Message Passing Interface (MPI) will be explored with future efforts to further refine the computational grid so that smaller cavities (on the order of $~10~\mu\mathrm{m}$) can be resolved.

\begin{figure}
\centerline{\includegraphics[width=0.43\textwidth]{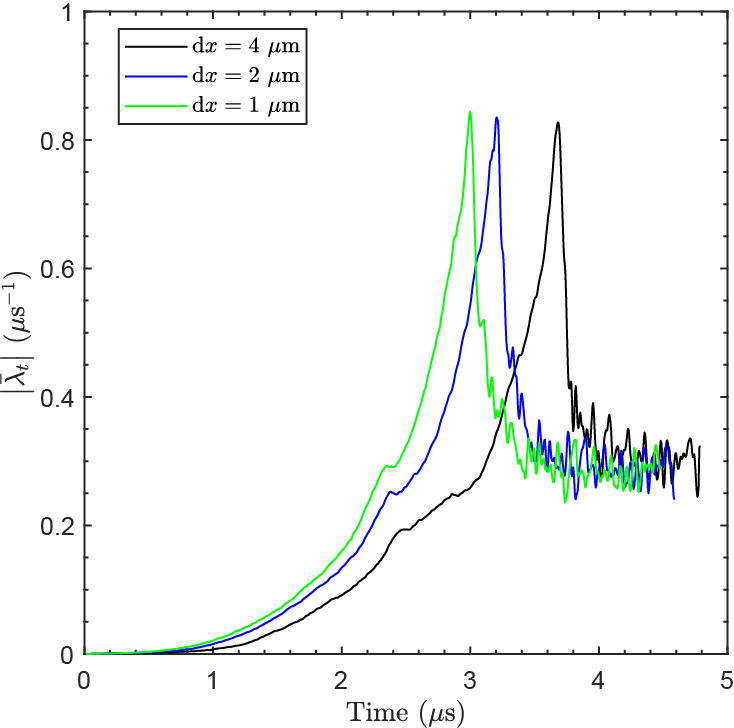}}
		\caption{Numerical convergence test: The normalized global reaction rate $\left| \bar{\lambda}_t \right|$ as a function of time for various computational grid sizes for the case with randomly distributed cavities of $\phi=5\%$ and $d_\mathrm{c}=100~\mu\mathrm{m}$ subjected to an input shock pressure of $7.81~\mathrm{GPa}$.}
	\label{FigA2}
\end{figure}

\begin{table}[ht]
\begin{center}
\caption{Numerical convergence tests for the case with randomly distributed cavities of $\phi=5\%$ and $d_\mathrm{c}=100~\mu\mathrm{m}$ subjected to an input shock pressure of $7.81~\mathrm{GPa}$.}
\label{Table_A2}
\begin{tabular}{| c | c | c | c |}
\hline
$\mathrm{d}x$ & $4~\mu\mathrm{m}$ & $2~\mu\mathrm{m}$ & $1~\mu\mathrm{m}$ \\
\hline
Overtake time & $3.68~\mu\mathrm{s}$ & $3.21~\mu\mathrm{s}$ & $2.99~\mu\mathrm{s}$\\
\hline
\end{tabular}
\end{center} 
\end{table}

\section{Statistical convergence study}
\label{append2}

\begin{figure}
\centerline{\includegraphics[width=0.43\textwidth]{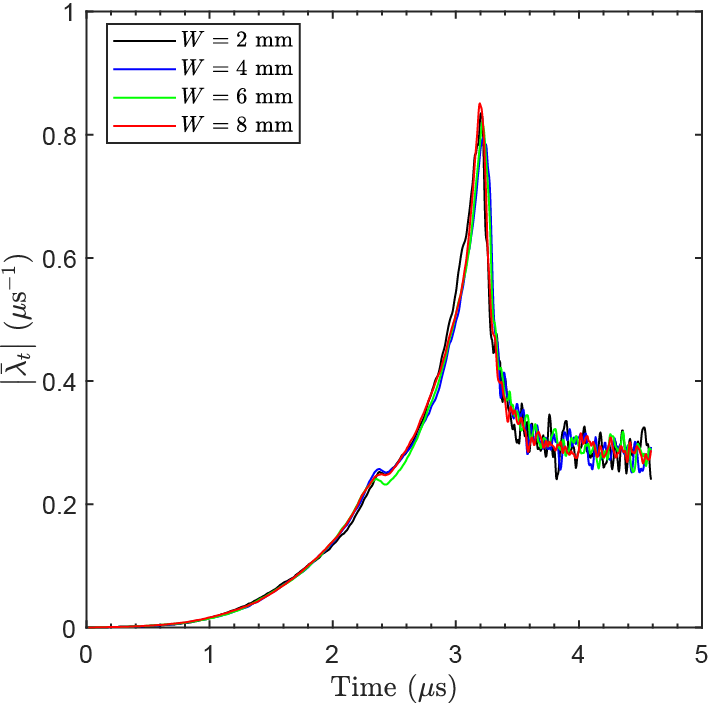}}
		\caption{Statistical convergence test: The normalized global reaction rate $\left| \bar{\lambda}_t \right|$ as a function of time at various computational domain widths $W$ (in $y$-direction) for the case with randomly distributed cavities of $\phi=5\%$ and $d_\mathrm{c}=100~\mu\mathrm{m}$ subjected to an input shock pressure of $7.81~\mathrm{GPa}$.}
	\label{FigA3}
\end{figure}

For the cases with randomly distributed cavities, it is of importance to verify whether the results, e.g., the detonation overtake time, are statistically converged so that further increasing the size of the random distribution would not significantly alter the results. To this end, a set of testing simulations has been carried out for the scenario with randomly distributed cavities of $\phi=5\%$ and $d_\mathrm{c}=100~\mu\mathrm{m}$ subjected to an input shock pressure of $7.81~\mathrm{GPa}$ at various widths ($W$) in the $y$-direction of the computational domain. The resulting temporal profiles of $\left| \bar{\lambda}_t \right|$ have been compared in Fig.~\ref{FigA3}. Note that, in order to extend the domain width beyond $W=2~\mathrm{mm}$ with a domain length of $24~\mathrm{mm}$, the numerical resolution of these simulations is chosen to be $\mathrm{d}x=2~\mu\mathrm{m}$. As shown in this figure, the profiles resulting from the simulations with $W=2$, $4$, $6$, and $8~\mathrm{mm}$ do not exhibit any significant difference in their overall behaviors and the detonation overtake time (where the value of $\left| \bar{\lambda}_t \right|$ peaks). Hence, in order to economize on the use of computing resource, the smallest domain width tested here, i.e., $W=2~\mathrm{mm}$, is selected for all of the simulations with randomly distributed cavities reported in this paper.  

{
\section{An alternative method to determine detonation overtake time}
\label{append3}

The trajectory of the leading shock front $x_\mathrm{s}(t)$ is tracked via finding the location where pressure first increases to $1.01$ times the initial pressure. A moving-averaged differentiation is performed to the obtained $x_\mathrm{s}(t)$ to calculate the time history of leading shock velocity $u_\mathrm{s}(t)$ and acceleration $a_\mathrm{s}(t)$ as follows, respectively,
\begin{equation}
    u_\mathrm{s}(t) = \frac{x_\mathrm{s}(t+\Delta \tau/2) - x_\mathrm{s}(t-\Delta \tau/2)}{\Delta \tau}
\end{equation}
and
\begin{equation}
    a_\mathrm{s}(t) = \frac{u_\mathrm{s}(t+\Delta \tau/2) - u_\mathrm{s}(t-\Delta \tau/2)}{\Delta \tau}
\end{equation}
where $\Delta \tau$ is the width of the time window over which the moving-averaged derivative is calculated. In order to filter high-frequency noises, $\Delta \tau = 1.5~\mathrm{\mu s}$ is chosen to obtain the data of shock acceleration reported in this Appendix.

The shock acceleration histories (thin orange curves against the right $y$-axis) obtained for the cases with neat NM and NM mixed with (b) regularly and (c) randomly distributed cavities of $\phi=8\%$ and $d_\mathrm{c}=100~\mu\mathrm{m}$ are compared with the corresponding histories of the overall reaction rate (thick blue curves against the left $y$-axis) in Fig.~\ref{FigA4}. The time at which the shock acceleration reaches its maximum closely agrees with the time of maximum $\left| \bar{\lambda}_t \right|$. Since the exact measurement of the time of maximum $a_\mathrm{s}$ depends on the time window width $\Delta \tau$ for moving-averaged differentiation (otherwise, influenced by high-frequency fluctuations in the velocity and acceleration histories), it is more convenient and accurate to use the approach of maximum overall reaction rate to determine detonation overtake time for the simulation results in this study.

\begin{figure}
\centerline{\includegraphics[width=0.47\textwidth]{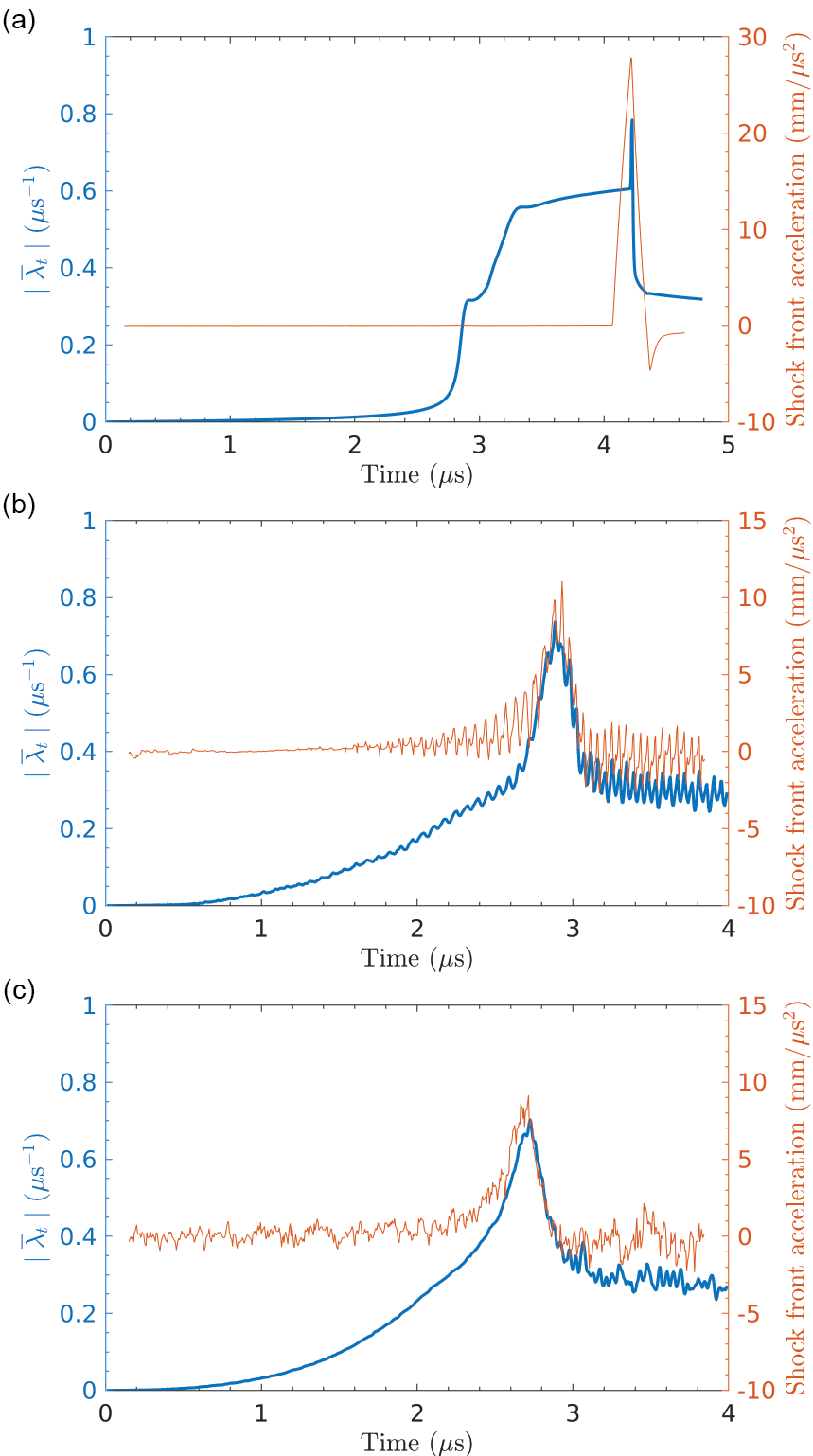}}
		\caption{The overall reaction rate $\left| \bar{\lambda}_t \right|$ (left $y$-axis) and leading shock acceleration (right $y$-axis) as functions of time for the case of (a) neat NM and the cases of NM mixed with (b) regularly and (c) randomly distributed cavities of $\phi=8\%$ and $d_\mathrm{c}=100~\mu\mathrm{m}$ subjected to an incident shock pressure of $7.81~\mathrm{GPa}$.}
	\label{FigA4}
\end{figure}
}

\bibliography{detonation}

\end{document}